\documentclass[pra,twocolumn,showpacs,amsfonts,amsmath,floatfix
]{revtex4-1}

\usepackage{graphicx}
\usepackage[usenames,dvipsnames,svgnames,table]{xcolor}
\usepackage[caption=false]{subfig}
\usepackage{dcolumn}
\usepackage{bm}

\newcommand{\be}{\begin{equation}}
\newcommand{\ee}{\end{equation}}
\newcommand{\ba}[1]{\begin{array}{#1}}
\newcommand{\ea}{\end{array}}
\newcommand{\bal}{\begin{align}}
\newcommand{\eal}{\end{align}}
\newcommand{\baln}{\begin{align*}}
\newcommand{\ealn}{\end{align*}}
\newcommand{\lr}[1]{\left( #1 \right)}
\newcommand{\ddd}{\mathrm{d}}

\begin{document}
\preprint{APS/123-QED}

\title{Versatile engineering of multimode squeezed states by optimizing the pump spectral profile in spontaneous parametric down-conversion}

\author{Francesco Arzani}
\email{francesco.arzani@lkb.upmc.fr}
\author{Claude Fabre}
\author{Nicolas Treps}

\affiliation{Laboratoire Kastler Brossel, UPMC-Sorbonne Universit\' e, CNRS, ENS-PSL Research University, Coll\` ege de France, 4  place  Jussieu,  75252  Paris,  France
}

\date{\today}

\begin{abstract}
We study the quantum correlations induced by spontaneous parametric down-conversion (SPDC) of a frequency comb. We derive a theoretical method to find the output state corresponding to a pump with an arbitrary spectral profile. After applying it to the relevant example of a spectrally chirped pump, we run an optimization algorithm to numerically find the pump profiles maximizing some target functions. These include the number of independently squeezed modes and the variances of nullifiers defining cluster states used in many continuous-variable quantum information protocols. To assess the advantages of pump-shaping in real experiments we take into account the physical limitations of the pulse shaper.

\end{abstract}

\maketitle

\section{Introduction}

Quantum Information Processing using Continuous Variables (CV)  of the electromagnetic field requires to efficiently generate various kinds of nonclassical states of light, such as squeezed or EPR entangled states. In addition, in CV quantum optics, it is relatively easy to scale up the dimensionality of the quantum system by increasing the number of modes of the field on which the quantum state spans. One of the most efficient and widely used techniques to produce in a deterministic way such multimode non-classical states of light is to use parametric down conversion in a $\chi^{\lr{2}}$ crystal. The quantum correlated, or "twin", photons  are then produced in different pairs of EPR entangled modes, the number of which depends on the characteristics  of the crystal and of the pump beam. These modes can be either spatial~\cite{lugiato1993spatial} or time/frequency~\cite{mosley} modes. In the case of frequency modes, which will be more specifically studied in this paper, the considered system is nothing else than the extension to the quantum domain of the classical WDM (Wavelength Division Multiplexing) technique, which has been so successful to increase by a large factor the performances of classical communications and information processing.
 
If one uses a single frequency pump, the different entangled signal and idler pairs modes, having frequencies symmetrically disposed with respect to the half pump frequency, are independent from each other. It is no longer the case when the pump spectrum contains more than one frequency. A bi-frequency pump already yields a quantum state with a complex structure of correlations~\cite{menicucci2008one,wang2014weaving,chen2014experimental} ; A mode-locked pump~\cite{roslund2014wavelength, de2014full, cai2017multimode}, producing trains of ultrashort pulses, i.e. a  frequency comb having millions of teeth, produces an even richer state in terms of quantum correlations and multipartite entanglement~\cite{gerke2015full,gerke2016multipartite}. Both systems are very promising sources of entangled \emph{cluster states}, a basic tool in the recently rising domain of CV Measurement-Based Quantum Computing (MBQC)~\cite{menicucci2006universal,gu2009quantum}.
 
The problem of optimizing the shape of the pump beam for a most efficient generation of a specific multimode quantum state of light is then of paramount importance for Quantum Computing applications. In the present paper we tackle the problem by the use of an algorithmic approach, having in view the optimized generation of specific cluster states, and in mind the experimental way to shape the pump, which consists in modifying the pump laser spectrum, both in phase and amplitude, using \emph{pulse shapers} based on the use of Spatial Light Modulators. We stress that such a setup allows to tailor many different pump shapes with no hardware modification to the experiment. Moreover, exploiting all the degrees of freedom provides great flexibility compared to engineering the phase-matching conditions or simply the width of the pump. The latter route has been explored before, often for Gaussian pulses only with at most quadratic spectral phase, especially in connection to the heralded production of single photons~\cite{URen1, *URen2, *tangle} or Fock states~\cite{optFock}. The focus of most earlier works on the subject was on the purity and entanglement of the signal and idler photons, which could be engineered to some extent tuning few parameters. This simplifications allowed to treat the problem analytically. The degree of control on the output state was correspondingly low and would not allow, for example, the optimization of a specific CV cluster state.

In the CV regime, which we are concerned with, the system is characterized by the quantum fluctuations in each mode. In this context, an analytic approach to general pump spectra with no spectral phase was developed for both spatial and temporal modes in~\cite{patera2012quantum}. However, the resulting theoretical profiles were not achievable with realistic experimental configurations. We show here that numerical optimizations can be fruitfully used to find the pump profiles producing multimode squeezed states with the properties needed for many different protocols. We also show that the numerical routines can be modified to take into account the physical limitations of a realistic pulse-shaper, ensuring that the optimized profiles are also experimentally realizable.
 
The article is organized as follows: in the next section we introduce mathematical tools based on Autonne-Takagi and Bloch-Messiah factorizations, that allow us to derive the covariance matrix of the output state in the frequency-mode basis for a pump field with an arbitrary spectral profile. We argue that these techniques are more suited to study type I collinear down-conversion, in which signal and idler photons are indistinguishable, than the singular value decomposition commonly used to treat non-degenerate down-conversion~\cite{walmsleySVD}. As an example we apply this formalism to the case of a spectrally chirped pump. In Sec.~\ref{sec:noise} we detail how the noise properties of an arbitrary set of temporal modes can be computed from the covariance matrix in the frequency basis. We introduce CV cluster states and explain how to compute their nullifiers for a given pump profile and a given set of modes. In Sec.~\ref{sec:optimize} we apply the described methods to the problem of the numerical optimization of the pump spectrum for several tasks: first we concentrate on the gains of the parametric down conversion. We then optimize the pump to produce cluster states whose nodes coincide with specific temporal modes having non-overlapping spectra. In Sec.~\ref{sec:conclusions} we summarize our results and comment on the generality of our approach, outlining how it may be fruitful for tackling problems beyond those detailed in the present work.

\section{Deriving the output state from the pump spectral profile}

We study type I spatially degenerate parametric down-conversion of an electric field containing a discrete set of $N$ frequencies inside a bulk crystal with $\chi ^{\lr{2}}$ nonlinearity. These frequencies may be, for example, the teeth of a frequency comb, or they may come from the discretization of a continuous broad band field, but our treatment can be applied to more general distributions of frequencies. Denoting by $a _j$ the annihilation operator at the frequency $\omega _j$, the quantum description of the process is given by the following hamiltonian in the pump interaction picture~\cite{patera2012quantum}\begin{equation} \label{eq:hamiltonian}
H_I = i \hbar \frac{ \eta } { 2 } \sum  _{\substack{j = 1 \\ k = 1}} ^N \mathcal{L}_{jk} a ^\dagger _j   a ^\dagger _k + \mathrm{h.c}
\end{equation} where $\mathrm{h.c.}$ means Hermitian conjugate. The real constant $\eta$ depends on the single-photon energy, the nonlinear susceptibility, the intensity and the geometry of the pump field. We assume that the spectrum is narrow enough so that photons at all the relevant frequencies have approximately the same energy. We consider only one spatial mode and only one polarization for the pump. Due to the collinear Type I design, the signal will also be in a single spatial mode and its polarization will be orthogonal to that of the pump. Frequency is then the only relevant degree of freedom. The coupling matrix $\mathcal{L}$ is the well known joint spectral distribution and is given by \begin{equation} \label{eq:LL}
\mathcal{L}_{jk} = \mathrm{sinc}\lr{\phi \lr{\omega_j,\omega_k}} \alpha\lr{\omega_j+\omega_k}.
\end{equation} The first factor is the phase matching function, with $\mathrm{sinc}\lr{x} = \sin \lr{x} /x$, and $\phi$ the phase mismatch angle \begin{equation} \label{eq:phaseMismatch}
\phi \lr{\omega_j,\omega_k} = \lr{k_p\lr{\omega_j+\omega_k} - k_s\lr{\omega_j} - k_s\lr{\omega_k} }\frac{l}{2}
\end{equation} $k_p\lr{\omega_j+\omega_k}$ being the wave number of the pump field at frequency $\omega_j+\omega_k$, $k_s\lr{\omega_k}$, the wave number of the signal field at frequency $\omega_k$ and $l$ denoting the length of the crystal. The dispersion relations for pump and signal fields can be computed using Sellmeier's equations (see Appendix~\ref{app:phiMBiBO}). The second factor in Eq.~(\ref{eq:LL}) is the complex spectral amplitude of the pump field. Earlier theoretical~\cite{valcarcel} and experimental~\cite{chen2014experimental,roslund2014wavelength} works have proven that hamiltonian (\ref{eq:hamiltonian}) accurately describes the physics of SPDC in the squeezing regime we will consider.

The joint spectral distribution has been widely studied in the specific case in which $\alpha\lr{\omega_j+\omega_k}$ is real for any $j$, $k$~\cite{patera2012quantum,walmsleySVD,temporalModes}, namely when the pump has no spectral phase up to a global phase factor. Since by construction $\mathcal{L}$ is symmetric, if the pump has no spectral phase $\mathcal{L}$ can be diagonalized with an orthogonal matrix, leading to uncoupled modes which are independently squeezed~\cite{patera2012quantum}. A slighly more sophisticated treatment is required to include pump shapes having arbitrary spectral phases. Examples of non trivial spectral phases can be met in fairly common situations, for example in the presence of a quadratic phase (spectral chirp). Two different approaches are possible: either diagonalizing the joint spectral distribution by congruence~\cite{autonne1915matrices, takagi1924algebraic,siegel} or applying the Bloch-Messiah decomposition~\cite{pramana,braunsteinSqueezing} to the symplectic transformation corresponding to a finite-time evolution of the system under the effective hamiltonian of the field inside the crystal. Diagonalization of a complex symmetric matrix by a congruence transformation through a unitary matrix is also known in the literature as Autonne-Takagi factorization or symmetric singular value decomposition. BM decomposition is another special case of singular value decomposition for symplectic matrices and is also known in the literature as Euler decomposition for symplectic matrices. We shall now detail both approaches and show how they allow to find modes of the electric field whose evolution is decoupled inside the crystal.

\subsection{Autonne-Takagi factorization \label{sec:takagi}}

Every complex symmetric matrix can be diagonalized by a congruence transformation with a unitary matrix~\footnote{Numerical routines for Autonne-Takagi factorization can be found in~\cite{hahn2006routines,takagiSVD}.}. This means that for any $\mathcal{L}$ in Eq.~(\ref{eq:LL}) one can find a unitary matrix $V$ such that \begin{equation} \label{eq:tak}
V\mathcal{L}V^T = \Lambda
\end{equation} with $\Lambda$ a diagonal matrix with real, non-negative entries. Suppose such matrix $V$ is known for a given $\mathcal{L}$, then one can define the vector of annihilation operators \begin{equation}
\vec{b} \equiv V^\dagger \vec{a}
\end{equation} with $\vec{a} = \lr{a_1, a_2, ... , a_N}^T$, each of which is a linear superposition of the single-frequency annihilation operators. Since $V$ is unitary, the operators $\vec{b}$ correspond to a set of orthonormal modes whose spectral profile is given by the rows of $V$. Substituting in Eq.~(\ref{eq:hamiltonian}) and using Eq.~(\ref{eq:tak}) one finds \begin{equation}\label{eq:takHam}
H_I = i \hbar \frac{ \eta } { 2 }\sum_k \Lambda_{kk}\lr{b^\dagger _k}^2 + \mathrm{h.c.}
\end{equation} showing that the modes $b_k$ evolve independently, each according to a squeezing hamiltonian.  These modes are referred to as \emph{supermodes} in the literature. The singular values $\Lambda_{kk}$ (multiplied by the parameter $\eta$) correspond to the gains of the downconversion process.

Note that having the same matrix $V$ on both sides of $\mathcal{L}$ in Eq.~(\ref{eq:tak}) was crucial to find the same decoupled modes for signal and idler (and thus a single creation operator $b_k ^\dagger$ for each $k$ in Eq.~(\ref{eq:takHam})). Ordinary (non-symmetric) singular-values decomposition would in general lead to different mode bases for the two. This is not a concern when treating non-degenerate SPDC in either polarization or spatial mode, since the signal and idler photons are distinguishable. However, for the problem at hand, the parametric interaction is more naturally described in terms of independent single-mode squeezers. Using singular-values decomposition, as opposed to Autonne-Takagi factorization, would require additional steps to achieve this.

\subsection{Bloch-Messiah decomposition \label{sec:BM}}

The previous approach solved the problem of finding the supermodes and the relative gains directly from the hamiltonian, which describes the differential evolution of the system. Although leading to the same physical results, it is sometimes more practical to work with the input-output relations corresponding to the evolution of the system for a finite time or its propagation over a finite crystal length. The main advantage is that fom this approach it is straighforward to derive the covariance matrix of the output state, encoding its noise properties. This is described in the following. 

Consider the equations of motion for the annihilation operators in the Heisenberg picture\begin{equation}
\frac{\ddd}{\ddd t } \vec{a}  = \frac{i}{\hbar} \left[ H_I , \vec{a}  \right] =  \eta \mathcal{L} \vec{a}^\dagger .
\end{equation} Complementing this set of equations with their adjoint one has \begin{equation}
\frac{\ddd}{\ddd t } \lr{\ba{c} \vec{a} \\ \vec{a} ^\dagger \ea} =  \eta \tilde{\mathcal{L} } \lr{\ba{c} \vec{a} \\ \vec{a} ^\dagger \ea}
\label{eq:annDiffEv}\end{equation} where \begin{equation}
\tilde{\mathcal{L}} =\lr{ \begin{array}{cc} 0 & \mathcal{L} \\ \mathcal{L}^* & 0\end{array} }.
\end{equation} Eq.~(\ref{eq:annDiffEv}) is readily integrated for a finite time $t$ \begin{equation} \label{eq:Sc}
\lr{\ba{c} \vec{a}\lr{t} \\ \vec{a} ^\dagger \lr{t}\ea }=  \exp\lr{\eta \tilde{\mathcal{L} } t} \lr{\ba{c} \vec{a}\lr{0} \\ \vec{a} ^\dagger \lr{0} \ea}.
\end{equation} We define the amplitude and phase quadrature operators respectively as \begin{align}
q_j &= \frac{a + a^\dagger}{\sqrt{2}} \\
p_j &= \frac{a - a^\dagger}{ i \sqrt{2}} .
\end{align} Introducing the matrix \begin{equation}
\mathcal{C} = \frac{1}{\sqrt{2}} \left( \begin{array}{cc}
\mathbb{I} & i \mathbb{I}\\ \mathbb{I} & -i\mathbb{I} 
\end{array} \right)
\end{equation} the quadrature operators of the relevant frequency modes are written as \begin{equation}
\lr{ \ba{c} \vec{q}  \\ \vec{p} \ea } = \mathcal{C}^\dagger \lr{\ba{c} \vec{a} \\ \vec{a} ^\dagger \ea }.
\end{equation} With this convention the shot noise $\Delta ^2 _0$ is normalized to $1/2$. Combining this with Eq.~(\ref{eq:Sc}) we find the expression for the finite-time evolution of the quadrature operators of frequency modes inside the crystal \begin{equation} S = \mathcal{C}^\dagger\exp\lr{ \eta  \tilde{\mathcal{L} } t  }\mathcal{C} \end{equation} so that \begin{equation}
\lr{ \ba{c} \vec{q} \lr{t} \\ \vec{p} \lr{t} \ea }  = S \lr{ \ba{c} \vec{q} \lr{0} \\ \vec{p} \lr{0} \ea }
\end{equation}
the hamiltonian $H_I$ can being applied for a time $t= l/c $. Actually $S$ is a spatial propagator corresponding to the input-output relation for the fields before and after the crystal. Changing the factor $\eta$ can easily be achieved adjusting the pump power (as long as it stays in the low-gain or below threshold regime in a cavity setup, which is the domain in which $H_I$ can be derived in the form used here).

Since $H_I$ is quadratic in the annihilation and creation operators, $S$ is a symplectic matrix. The matrix $\mathcal{C}$ links it to its complex representation $S^{\lr{c}} = \exp ( \eta  \tilde{\mathcal{L} } t ) $, appearing in Eq.~(\ref{eq:Sc})~\cite{pramana}. We can apply the Bloch-Messiah decomposition and find a factorization~\cite{braunstein2005squeezing} \begin{equation}
S = R_1 K R_2 \label{eq:BMgen}
\end{equation} where $R_1$ and $R_2$ are both symplectic and orthogonal matrices and $K = \mathrm{diag}\left\{ e^{r_1}, e^{r_2}, ..., e^{r_N},   e^{-r_1}, e^{-r_2}, ..., e^{-r_N} \right\}$ is a squeezing matrix, namely a symplectic diagonal matrix. Its diagonal entries are the singular values of $S$. In our case, single-frequency modes are the input and output of the overall process, so $R_2 = R_1 ^{-1} = R_1 ^T$ and $S$ is symmetric~ \footnote{Bloch-Messiah decomposition has a simple physical intrepretation. In an interferometric picture in which modes are spatially separated, $R_1$, $R_2$ represent linear optical networks, and $R_2$ can simply be omitted when the input state is vacuum. A rigorous treatment requires some caution when dealing with more general modes, such as the combinations of single-frequency modes considered here. $R_2$ must then be taken into account to relate the input modes, defined by the experimenter, to the squeezed modes.}. The spectral profiles of the supermodes are the rows of the unitary matrix $ U $ appearing in the complex representation of $R_1$~\cite{pramana} \begin{equation} R_1 ^{\lr{c}} \equiv \mathcal{C} R_1 \mathcal{C} ^\dagger = \mathrm{diag}\left\{ U, U^* \right\} .\end{equation} As we will see in the next section, the supermodes found in this way are the same as those obtained through the Autonne-Takagi factorization.

In the hypothesis that the system was initially in the vacuum state, the covariance matrix in the frequency modes basis can also be computed from $S$ as~\cite{ferraro2005gaussian} \begin{equation}
\Gamma _\omega = \frac{1}{2} S S^T = \frac{1}{2} R_1 K^2 R_1 ^T  .
\end{equation} Note that it is not necessary to compute the Bloch-Messiah decomposition to get the covariance matrix from $S$.

\subsection{Relating the two approaches}

Given the Autonne-Takagi factorization of $\mathcal{L}$, it is straightforward to compute the Bloch-Messiah decomposition of $S$. In fact, defining \begin{align}
R_1 &=   \mathcal{C}^\dagger \mathrm{diag}\left\{ V^\dagger, V^T \right\} \mathcal{C} \\ R_2 &= R_1 ^T
\end{align} one finds \begin{align}
R_1 ^T S R_2 ^T &= R_1 ^\dagger S R_2 ^\dagger \\ &=  \mathcal{C}^\dagger \exp \left\{ \eta t \lr{\ba{cc} 0 & V \mathcal{L} V^T \\ V^* \mathcal{L}^* V^\dagger  & 0 \ea }   \right\} \mathcal{C} \\
 &=  \exp \left\{ \eta t \lr{\ba{cc} \Lambda & 0 \\ 0  & -\Lambda  \ea }   \right\} = K 
\end{align} where $K$ is the same as in Eq.~(\ref{eq:BMgen}) (up to permutations of the diagonal elements). The advantage of using Autonne-Takagi factorization is that it is numerically easier to compute with respect to Bloch-Messiah decomposition. The link between Autonne-Takagi factorization and Bloch-Messiah decomposition was also recently noted in~\cite{padova}.

\subsection{Numerical simulations \label{sec:numDet}}

Most of our results are obtained through numerical simulations, the details of which can be found in Appendix~\ref{app:numDet}. For our examples, we take the unshaped pump $\alpha ^{\lr{\mathrm{g}}} \lr{\omega}$ to be a Gaussian pulse of spectral width about $\Delta \lambda \approx 3.54$ nm full-width-half-maximum (FWHM)  centered around $\lambda_0 =397.5$ nm, which can be obtained by upconversion of a $10$ nm pulse FWHM, corresponding to a duration of about $100$ fs, centered around $795$ nm. We consider free-space setups and assume the nonlinearity is provided by bulk BIBO crystals of length between $0.5$ mm and $2$ mm, whose refractive indexes are computed using Sellmeyer's equations (see Appendix~\ref{app:phiMBiBO} for details). We denote the unshaped spectral profile by \begin{equation} \label{eq:refGauss}
\alpha ^{\lr{\mathrm{g}}} \lr{\omega} = \frac{1}{\sqrt{\sigma_\omega \sqrt{2\pi}}}\mathrm{e}^{-\frac{\lr{\omega - \omega _0}^2 }{4 \sigma_\omega ^2 }}
\end{equation} with $\omega_0 = 2\pi c / \lambda_0$ and $\sigma_\omega = \omega_0 ^2 \Delta \lambda / 4\pi c \sqrt{2\ln 2}$, $c$ being the speed of light in vacuum. 

In previous works considering a real pump with a Gaussian spectrum~\cite{patera2012quantum}, it was noted that the diagonalization of $\mathcal{L}$ leads to alternating signs in the gains, meaning that the supermodes are squeezed in alternating quadratures. This actually comes from imposing that the spectral profile of the supermodes is real, which is possible because the supermodes have a trivial spectral phase. An equivalent choice would be to define the supermodes to be all squeezed in the same quadrature, which amounts to multiplying the spectral amplitudes of half of the supermodes by $i$. In fact, multiplying a row of $V$ by $i$ in Eq.~(\ref{eq:tak}) flips the sign of the corresponding diagonal element in $\Lambda$ and rotates the squeezing direction by $\pi/2$ in phase space. Defining the modes such that the phase quadrature is always the squeezed one is more suited to handle the case in which the pump has a non-trivial spectral phase. The reason is that in this case supermodes may have non-trivial specral phases as well, as we shall see, so there is no simple criterion to choose which quadrature should be squeezed based on supermodes.

To keep close to an experimental scenario, we assume that the spectral profile of the pump is modified by a pulse shaper, which can be built with a spatial light modulator in a 4-f configuration~\cite{pulseShaper}. We model this as a device with a finite number of degrees of freedom, parametrized by the real vector $\vec{u}$, corresponding to spectral amplitude and phase at given frequencies. The resulting pump profile $\alpha ^{\lr{\vec{u}}} \lr{\omega}$ is found multiplying the unshaped amplitude by a transfer function $\mathcal{I} ^{\lr{\vec{u}}}  \lr{\omega}$ interpolating these parameters \begin{equation}   
\alpha ^{\lr{\vec{u}}} \lr{\omega} = \alpha ^{\lr{\mathrm{g}}} \lr{\omega} \mathcal{I} ^{\lr{\vec{u}}}  \lr{\omega}.
\end{equation} More details can be found in Appendix~\ref{sec:shaper}.

It is worth clarifying how we derive physical values for the squeezing of the supermodes. These are proportional to the factor $\eta$ in the Hamiltonian of Eq.~(\ref{eq:hamiltonian}), which is generally difficult to compute accurately from first principles. For our purposes, it will be more convenient to adjust it so that the squeezing of the first supermode approximately matches the experimentally measured value. Once the highest squeezing is fixed, the ratio between the squeezing parameters of the supermodes is the same for any pump power below threshold~\cite{patera2012quantum}.

\subsection{An example: Chirped pump \label{sec:chirp}}

\begin{figure}
    \subfloat[]{
        \centering
        \label{subFig:firstChirp}
        \includegraphics[width=0.49\textwidth]{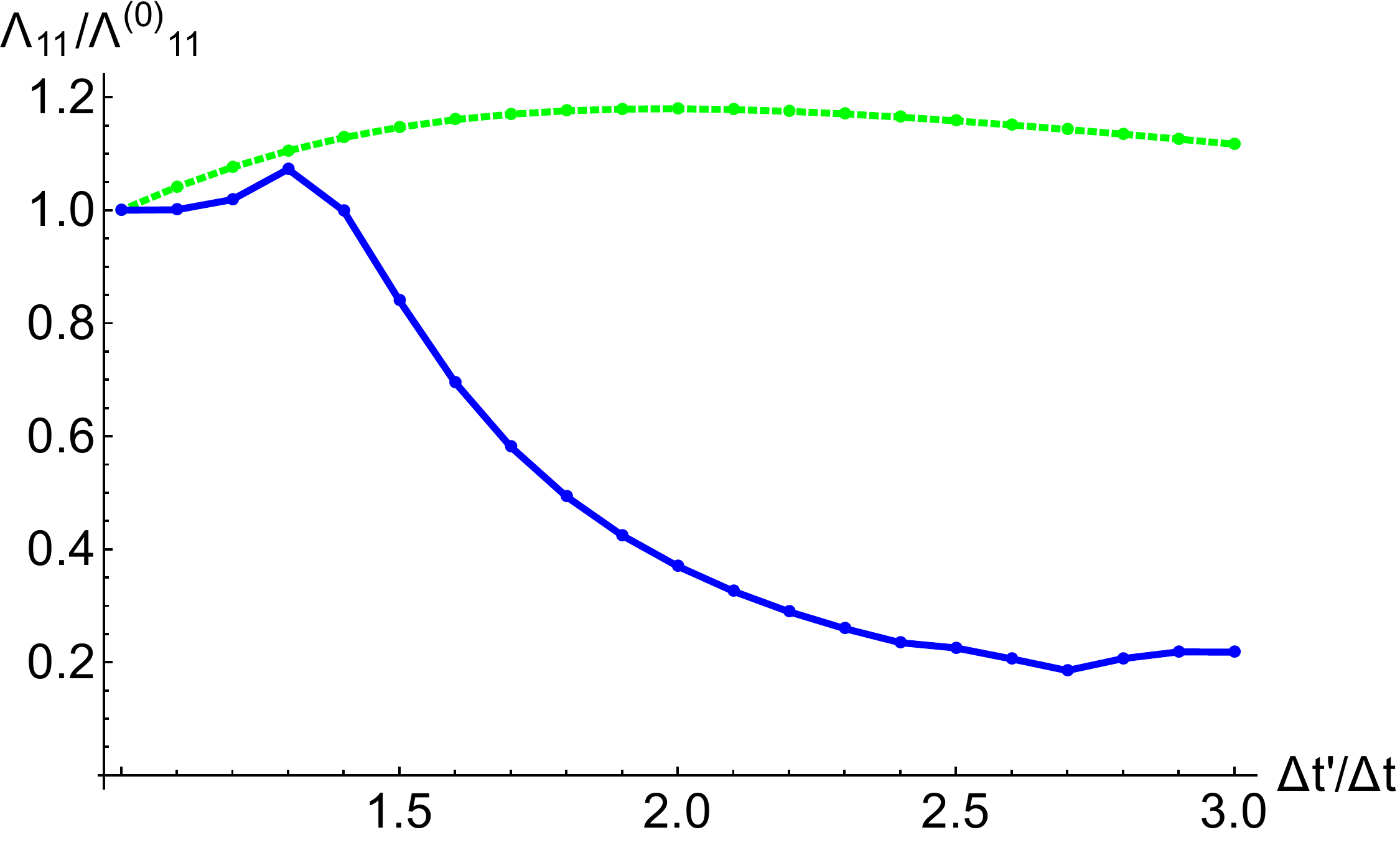}
    }\\
    \subfloat[]{
        \centering
        \label{subFig:chirpDistr}
        \includegraphics[width=0.49\textwidth]{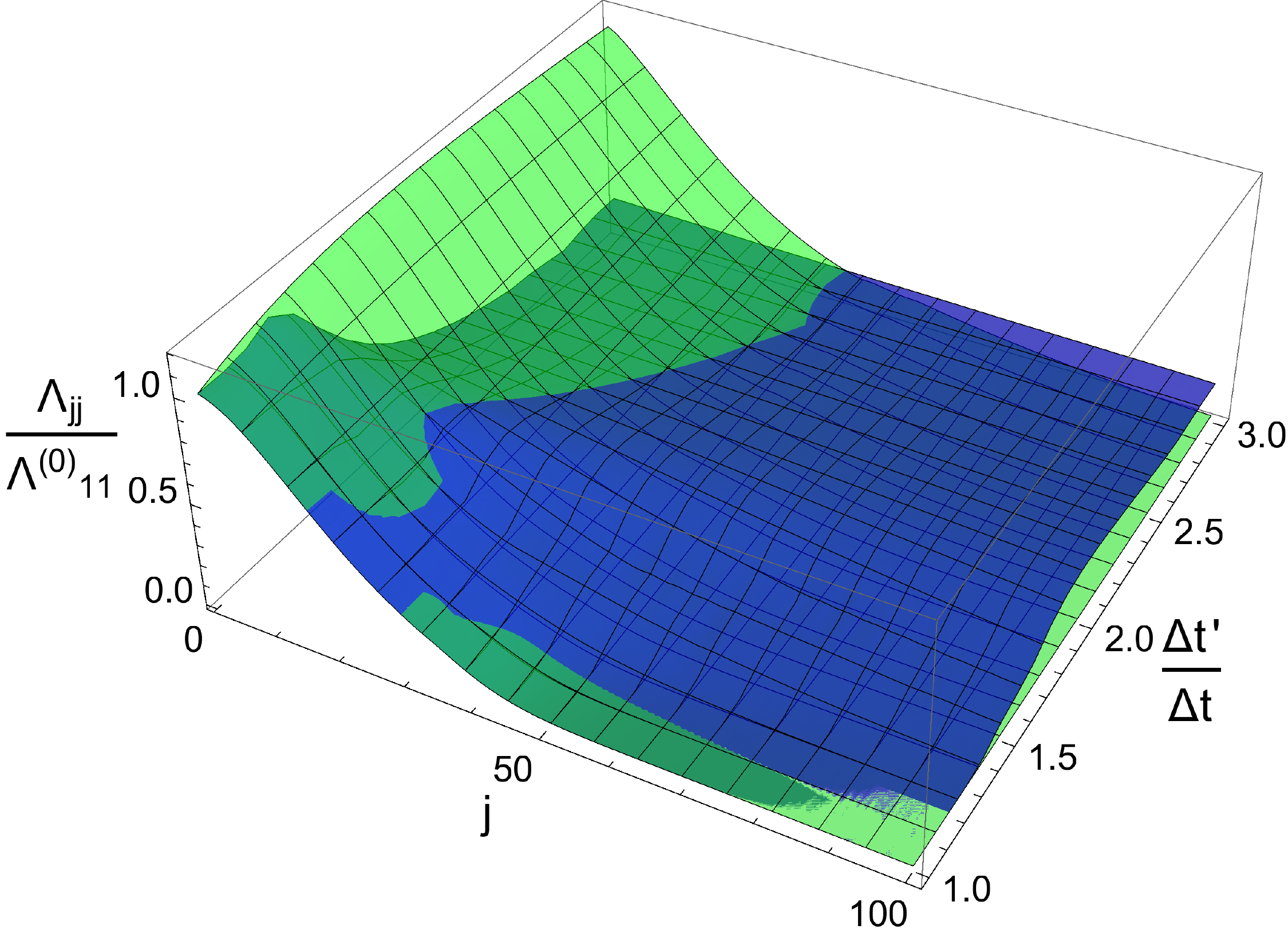}
    }\\
    \subfloat[]{
        \centering
        \label{subfig:chSM}
        \includegraphics[width=0.49\textwidth]{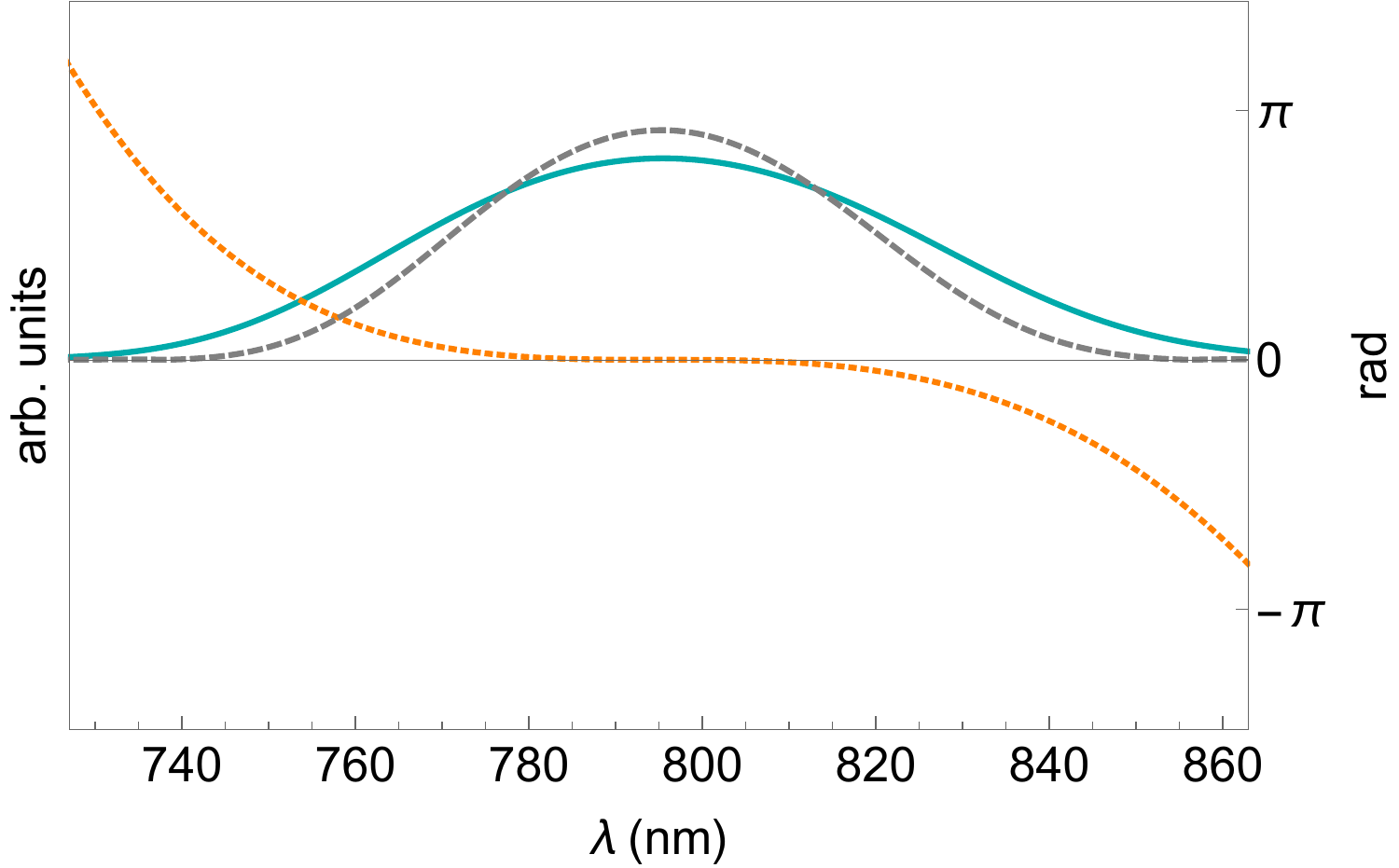}
    }
    
    \caption{\label{fig:chirpSqz}(Color online) Comparison between the effect of the quadratic phase and simply increasing the pulse duration. (a) Largest parametric gain for a chirped (dotted green curve) and non-chirped (solid blue curve) as a function of the increase in pulse duration $\Delta t ' / \Delta t$. The gains are all divided by the largest parametric gain for $\phi_{2} = 0$ and $\Delta t ' / \Delta t = 1$, denoted by $\Lambda^{\lr{0}}_{11}$. (b) First one hundred gains for increasing pulse duration for chirped (dark blue) and non-chirped (light green) pulses (normalized as in (a)). (c) Spectral amplitude (blue, solid line) and phase (orange, dotted line) of the first supermode obtained with $\Delta t ' = 2 \Delta t $ ($\phi _2 \approx 2700\ \mathrm{fs}^{-2}$ ) compared to the first supermode for $\phi_{2} = 0$ (gray, dashed line). } 
\end{figure}

As a first example of a pump with a non-trivial spectral phase we consider a gaussian pump with a quadratic spectral phase, namely a spectrally chirped pump of amplitude \begin{equation}
\alpha ^{\lr{\mathrm{ch}}}\lr{\omega} = \alpha ^{\lr{\mathrm{g}}} \lr{\omega}  e^{i\frac{\phi_2}{2} \lr{\omega - \omega_0} ^2 }
\end{equation} where $\phi_2$ is the quadratic phase. Spectral chirp is fairly common in experimental situations, often as an unwanted effect, so it is interesting to study its impact on the down-conversion process. The quadratic spectral phase implies that the pulse is no longer Fourier limited: the duration of the pulse increases while the spectrum remains constant. This makes the duration of the pulse a useful parameter to characterize the amount of chirp. If $\Delta t =  1/(2\sigma_\omega)$ is the duration of the un-chirped pulse~\footnote{With the convention $\Delta t = 1/(2\sigma_\omega) $, $\Delta t$ is also the standard deviation of the temporal envelope $ \tilde{\alpha} ^{\lr{\mathrm{g}}} \lr{t} = \lr{2\pi}^{-\frac{1}{2}}\int \ddd \omega \alpha ^{\lr{\mathrm{g}}} \lr{\omega} \exp \lr{i\omega t}$, where $\alpha ^{\lr{\mathrm{g}}} \lr{\omega}$ is the Gaussian spectral envelope defined in Eq.~(\ref{eq:refGauss})~\cite{thielThesis}. Namely $\Delta t ^2 = \int \ddd t \ t^2 \left| \tilde{\alpha} ^{\lr{\mathrm{g}}} \lr{t} \right|^2 $ } ($\phi_2 = 0$), the duration after chirp is~\cite{thielThesis} \begin{equation}
\Delta t ' = \Delta t \sqrt{ 1 + \lr{\frac{\phi _2}{2 \Delta t ^2 }}^2 }.
\end{equation} Studying the dependence of the output state it is then natural to ask how much modification is really due to the spectral phase and how much is just a consequence of the increased duration. It turns out that the two situations are very different, as can be seen from the plots in Fig.~\ref{fig:chirpSqz}. We compare, for the two cases, the largest parametric gain (Fig.~\ref{subFig:firstChirp}) as well as the first one hundred parametric gains (Fig.~\ref{subFig:chirpDistr}) as functions of $\Delta t ' / \Delta t $. The plots were obtained for a fixed energy in each pump pulse. We assume the downconversion of a pulse with $\Delta t \approx 30 $ fs takes place in a 0.5 mm BIBO crystal. All the gains are normalized to the highest gain for $\phi_2 = 0$ and $\Delta t ' / \Delta t = 1$. In both cases, the gain of the first supermode $\Lambda_{11}$ increases with $\Delta t '$  at first but then starts decreasing. However, the descent is steeper in the chirped case. Moreover, numerically we find that for increasing quadratic phase \begin{equation}
\Lambda_\mathrm{tot} = \sum_j \Lambda _{jj} ^2 = \mathrm{const.}
\end{equation} within machine precision~\footnote{This is actually easy to prove analytically~\cite{dimaChirp}. In fact $\sum_j \Lambda_{jj}^2 = \mathrm{Tr}\lr{\Lambda \Lambda^*}$ and using the symmetry of the matrix $\mathcal{L}$ one has $\sum_{jk} \mathcal{L}_{jk}\mathcal{L}_{jk}^* = \sum_{jk} |\mathcal{L}_{jk}|^2 $ which is manifestly independent of the quadratic phase.}, whereas $\Lambda_\mathrm{tot}$ monotonically increases for un-chirped pulses of longer duration. To get a physical picture of $\Lambda_\mathrm{tot}$, consider the perturbative expansion of the evolution for small times/pump power/nonlinearity. The $\Lambda_{jj}$ are then seen to be proportional to the probability amplitude for a pump photon to be converted into two photons in the supermode $j$. In fact, applying the evolution operator for a small time $\delta t$ to the vacuum one gets \begin{align}
U\lr{\delta t} \left|0\right>  &= \sum \limits _{l=0} ^{\infty} \frac{\lr{- i \delta tH_I}^l}{l!\hbar ^l} \left|0\right> \\ & = \lr{\mathbb{I} + \delta t \frac{ \eta } { 2 }\sum_k \Lambda_{kk}\lr{b^\dagger _k}^2  + \mathcal{O}\lr{\delta t^2}}\left|0\right> .
\end{align} The sum of $\Lambda_{jj} ^2$ is then proportional to the probability of converting a photon of the pump into two photons in any supermode within time $\delta t$. This can be interpreted as the conservation of the overall efficiency of the down-conversion process for increasing quadratic phase. On the other hand, it is clear that the process is not insensitive to the quadratic spectral phase: more signal modes are excited as the quadratic phase increases, while the highest gain for a single mode decreases. The overall efficiency increases for un-chirped pulses of longer duration, but the magnitude of the gains drops faster. As a consequence, for large $\Delta t ' $ the number of modes with approximately the same squeezing is higher for a chirped pump, as can be seen from Fig.~\ref{subFig:chirpDistr}. Chirp can be added easily in experiments at constant pump power, whereas changing the pulse duration generally involves losses. The former may then be more convenient.

Fig.~\ref{subfig:chSM} shows spectral amplitude and phase of the first supermode obtained for $\phi_2  \approx 2700 \  \mathrm{fs}^2$, the quadratic phase doubling the duration of the pulse. For the plot, we subtracted a linear term from the spectral phase, which only amounts to a temporal delay. Interestingly, the remaining spectral phase is not quadratic, as in the pump. Instead, it is well fitted by a cubic term \begin{equation}
\phi_{\mathrm{fit}}\lr{\omega} = e^{i\phi_3\lr{\omega - \omega_0/2}^3}.
\end{equation} The same cubic phase fits well the spectral phase of all the supermodes and is thus an important effect to take into account in experiments. The coefficient $\phi_3$ seems to have a non-trivial dependence on $\phi_2$. A systematic study of the effect of chirp is however beyond the scope of the present work and is left to future investigations. We only wish to highlight here that Autonne-Takagi factorization can be used to study pump fields with arbitrary spectral shapes and this can lead to the discovery of new and interesting features already in quite simple situations. 

\section{Noise properties of the output state \label{sec:noise}}

Here we introduce the formalism we will use to compute the relevant measurable quantities of the output state from the covariance matrix in the frequency basis.

\subsection{Noise of a set of modes}

The noise properties of any mode can be computed from the covariance matrix in the frequency basis $\Gamma _\omega$. Take the mode corresponding to the annihilation operator \begin{equation}
d = \sum _l v_l a_l \label{eq:detAnn}
\end{equation} where the $v_l$ are complex numbers satisfying $\sum _l \left| v_l \right| ^2 =1$. $v_l$ is the complex amplitude of the electric field mode at frequency $\omega_l$. The quadratures of $d$~\footnote{We will identify modes with their annihilation operator.} are given by \begin{align}
q^{\lr{d}} &=  \sum _l \lr{ \mathrm{Re}\lr{v_l} q_l - \mathrm{Im}\lr{v_l} p_l }\\
p^{\lr{d}}  &=  \sum _l \lr{ \mathrm{Im}\lr{v_l} q_l + \mathrm{Re}\lr{v_l} p_l }.
\end{align} Consider now a set of $M \leq N$ orthogonal modes related to the frequency modes by \begin{equation}
\vec{d} = D \vec{a} \label{eq:dModes}
\end{equation} where the matrix $D$ has $M\times N$ complex entries. The orthonormalization condition of the modes takes the form \begin{equation}
D D^\dagger  = \mathbb{I}_M .
\end{equation} The quadratures of modes $\vec{d}$ are then given by \begin{equation}
\lr{ \ba{c} \vec{q}^{\lr{d}} \\ \vec{p}^{\lr{d}}\ea }  = \lr{ \ba{cc} \mathrm{Re}\lr{D} & - \mathrm{Im}\lr{D} \\  \mathrm{Im}\lr{D} & \mathrm{Re}\lr{D}\ea } \lr{ \ba{c} \vec{q} \\ \vec{p} \ea } \equiv R_D \lr{ \ba{c} \vec{q}  \\ \vec{p} \ea } . \label{eq:RV}
\end{equation} The covariance matrix of the modes $\vec{d}$ is then obtained from that of frequency modes as \begin{equation}
\Gamma _d = R_D \Gamma _\omega R_D ^T. \label{eq:detCov}
\end{equation} When $M<N$, the transformation in Eq.~(\ref{eq:detCov}) can be understood as changing the modes to a basis of which $\vec{d}$ constitute the first $M$ elements and then discarding the remaining modes (which amounts to removing the corresponding rows and columns from the covariance matrix).

\subsection{Cluster states and nullifiers \label{sec:nulls}}

One of the main goals of our work is to exploit the methods outlined above in optimization routines to find the shape of the pump which is best suited to produce CV cluster states on a given set of modes. In order to do this, we recall that a CV cluster state is a multimode state which, in its ideal version, can be defined as the simultaneous eigenstate of a set of operators called nullifiers. If $G$ is the graph associated with the cluster state, which we will identify with its adjacency matrix, nullifiers can be written as \begin{equation}
\vec{\delta} = \vec{p}^{\lr{\mathrm{d}}} - G \vec{q}^{\lr{\mathrm{d}}}. \label{eq:nulls}
\end{equation} We assumed here that the nodes of the cluster correspond to the generic modes $\vec{d}$ of Eq.~(\ref{eq:dModes}). Although more general situations can be considered~\cite{graphCalc}, we will restrict to unit-weight cluster states. In this case $G_{jk} = 1$ if and only if modes $j$ and $k$ are nearest neighbours in the graph and all the other entries of $G$ are zero. Different conditions may be required to certify the experimental production of cluster states, but a basic one is that the noise of the nullifier operators lay below the vacuum noise. Standard homodyne detection techniques are sufficient to measure the quantum fluctuations of these operators, as explained in Appendix~\ref{app:nullsMeas}. 

\section{Pump optimization \label{sec:optimize}}

From the previous sections it should be clear that the relation between the spectral profile of the pump and the properties of the output state is far from trivial. As a consequence, it is generally very difficult to find an analytical form for the pump optimizing a given property of the output, such as the entanglement pattern of a given set of modes. Instead, one could run a numerical optimization algorithm to try and improve the desired quantities. The results obtained with this approach make the object of the following sections. For the optimization we used an evolutionary algorithm  developed in~\cite{jonOpt}, of which some details are given in Appendix~\ref{app:opt}.

\subsection{Squeezing spectrum}

\begin{figure*}
    \begin{minipage}{0.49\textwidth}
    \subfloat[]{
        \centering
        \label{subFig:spectra}
        \includegraphics[width=\textwidth]{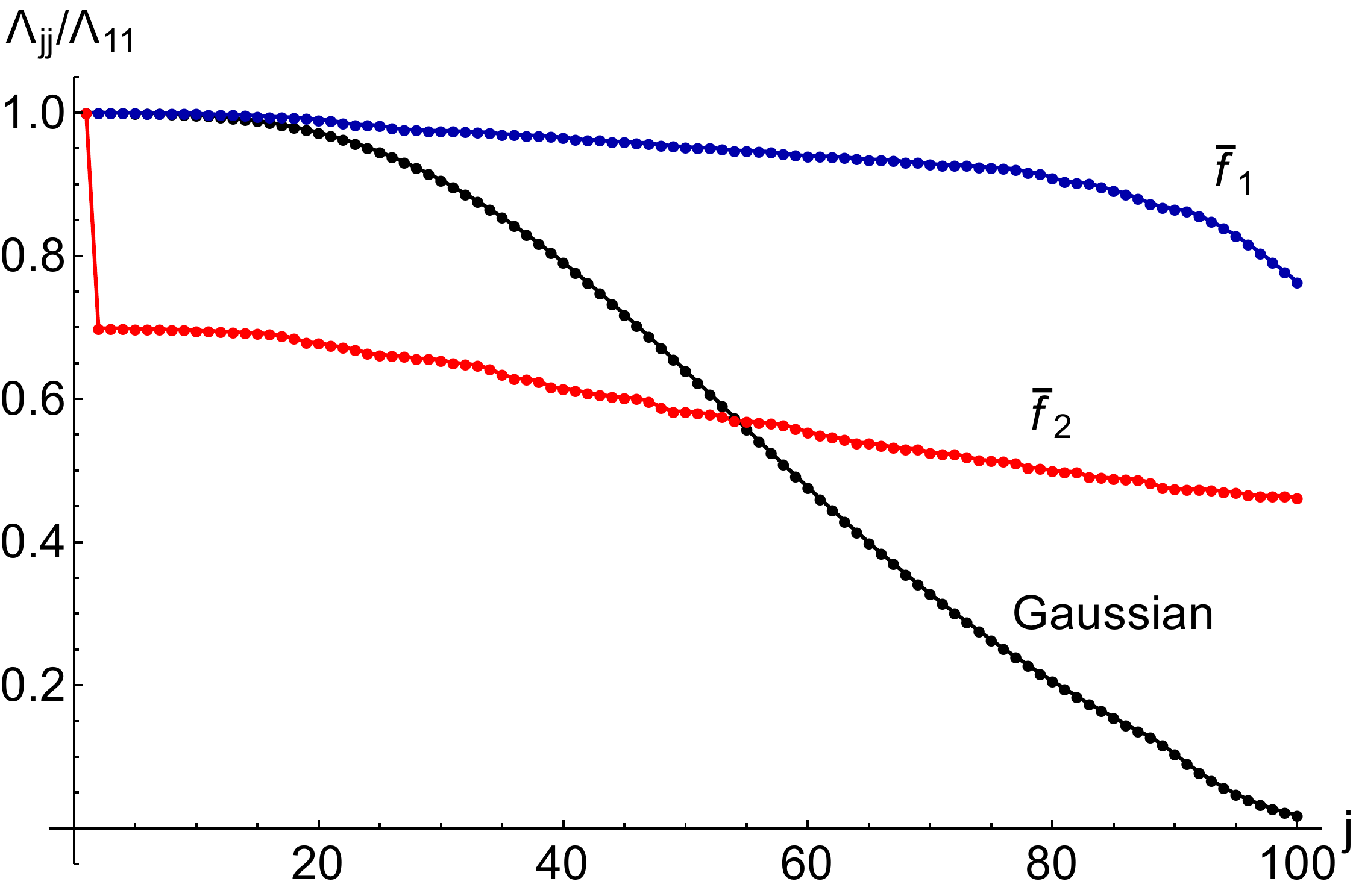}
    }
\end{minipage}    \subfloat[]{
    \begin{minipage}{0.49\textwidth}
        \centering
        \label{subFig:supermodes}
        \includegraphics[width=\textwidth]{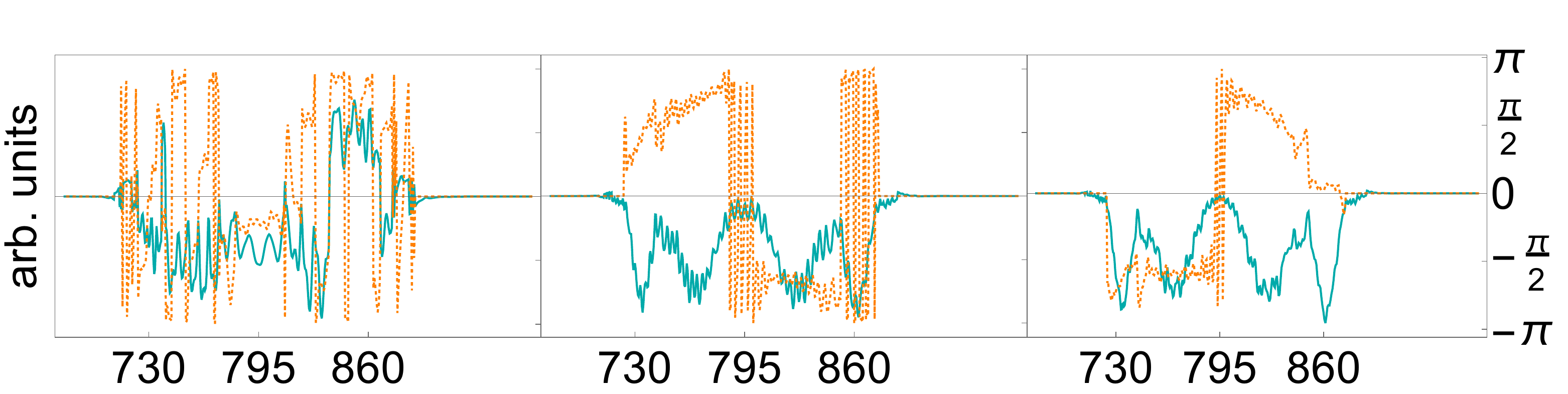}
        \includegraphics[width=\textwidth]{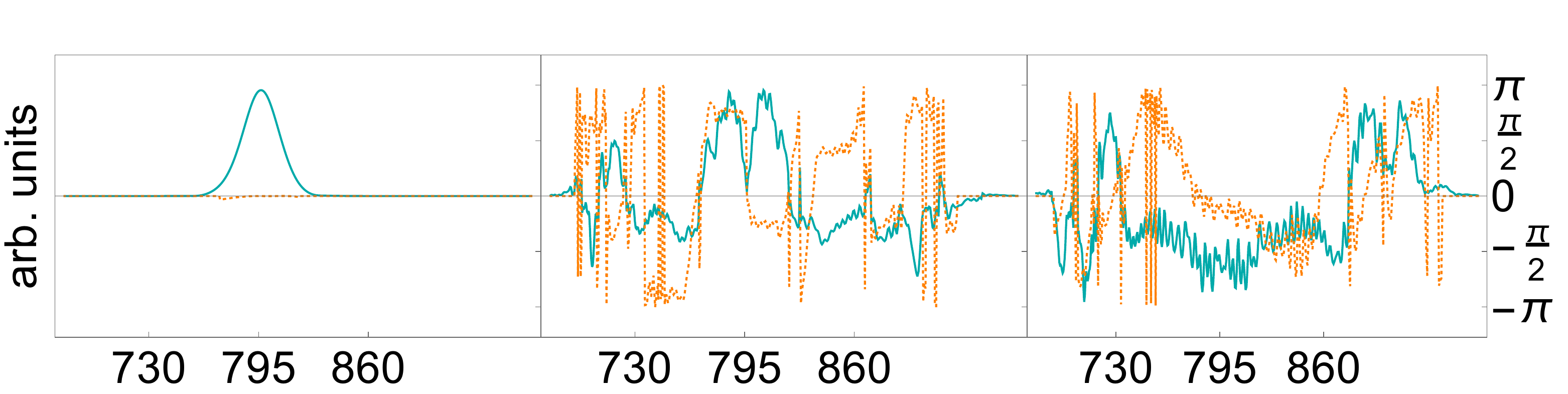}
    \end{minipage}
    }   %
    
    \subfloat[]{
        \centering
        \label{subFig:f1Pump}
        \includegraphics[width=0.49\textwidth]{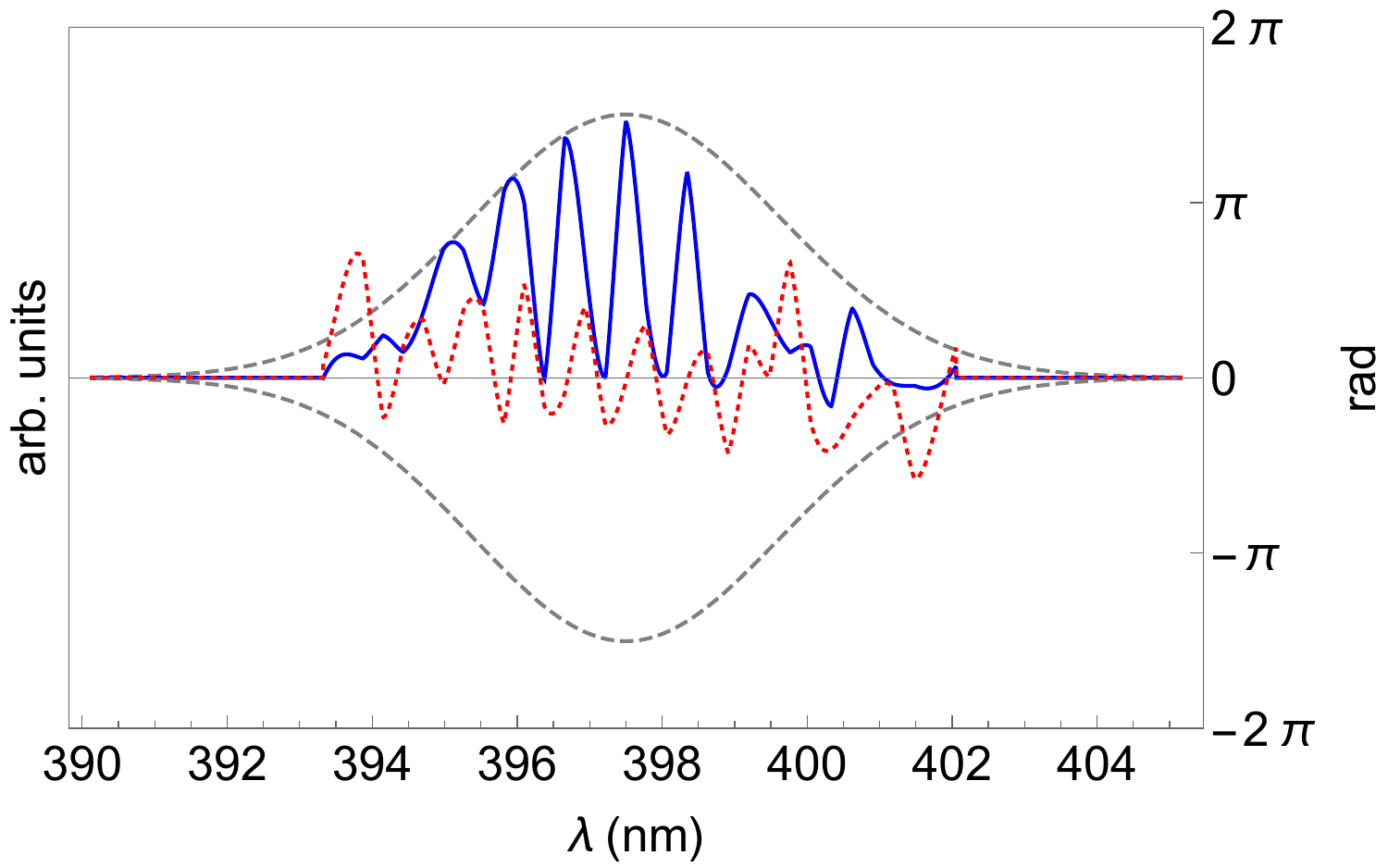}
    }
    \subfloat[]{
        \centering
        \label{subFig:f2Pump}
        \includegraphics[width=0.49\textwidth]{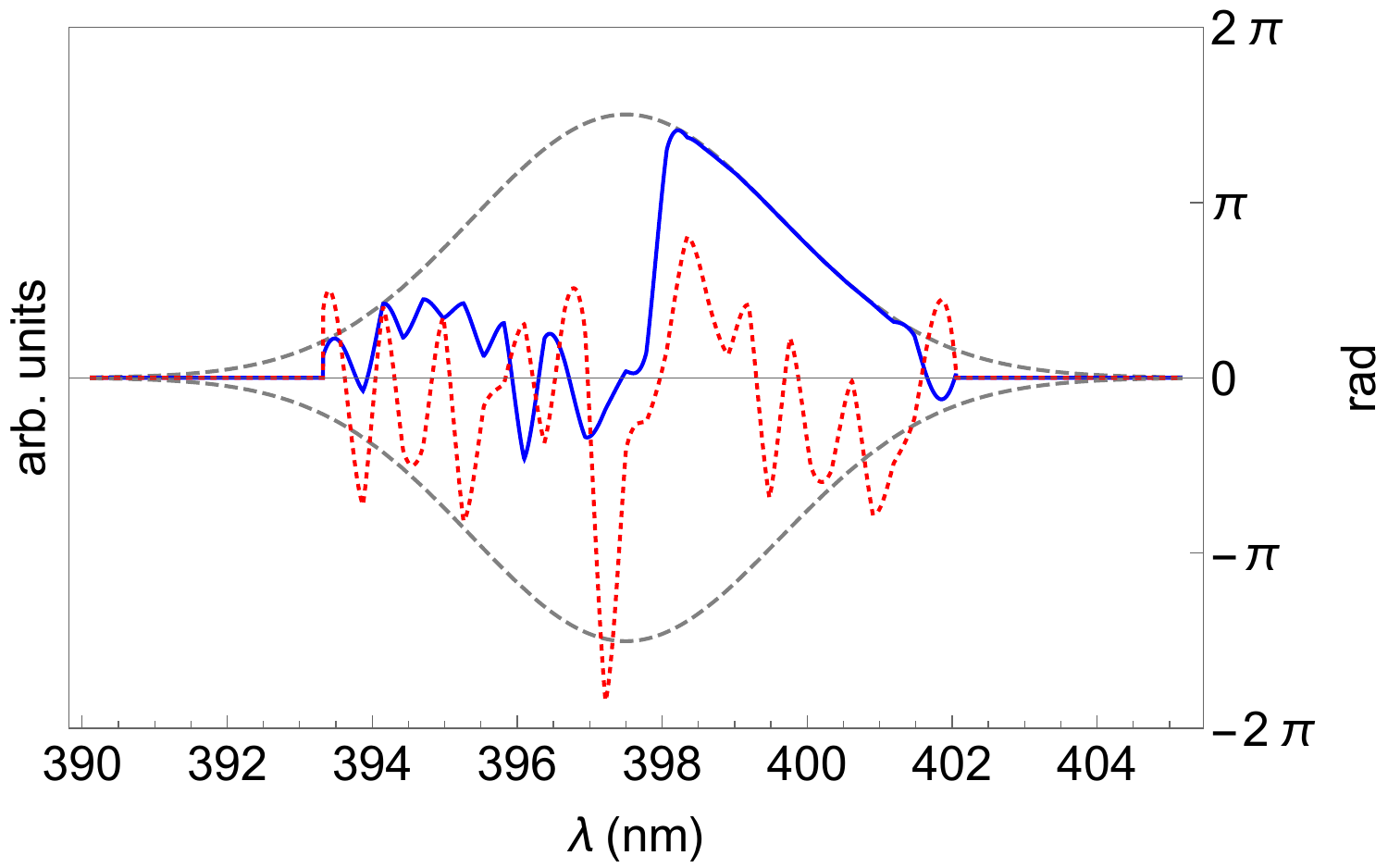}
    }
    \caption{\label{fig:optSpectra}(Color online) (a) The normalized gain distributions obtained for a Gaussian pump and after optimizing $\bar{f}_1$ in Eq.~(\ref{eq:f1}), with $a = 3 $, $x\lr{w} = 1/\lr{5w}^6$ and $k=100$ and $\bar{f}_2$ in Eq.~(\ref{eq:f2}), with $b = 1$, $y\lr{w} = 1/\lr{5w}^6$. (b) The first supermodes resulting from the pump optimizing $\bar{f}_1$ (top) and $\bar{f}_2$ (bottom). The solid blue line represents the amplitude, in arbitrary units, while the orange dashed line represents phase, in radiants (scale on the right). For clarity of representation we subtracted a linear phase of $260$, $812$ and $805$ fs from the supermodes arising from the optimization of $\bar{f}_1$ and of $275$, $275$ and $-390$ fs from the supermodes arising from the optimization of $\bar{f}_2$.  (c) and   (d)  show the pump profiles maximizing $\bar{f}_1$ and $\bar{f}_2$, respectively. The gray dashed line shows the original Gaussian, the solid blue line the optimal amplitude profile and the red dotted line the optimal phase.} 
\end{figure*}

We have already seen in Sec.~\ref{sec:chirp} that changing the spectral profile of the pump can impact the squeezing spectrum. We investigate here to which extent this can be used to enhance a given property. Specifically, we look for the spectral profiles that flatten the squeezing spectrum, equalizing the first $k$ gains, or separate the highest gain from the others, effectively concentrating more squeezing in the first supermode. For the first task, we run the optimization for the fitness function \begin{equation}
f_1 \lr{\vec{u}} = \frac{1}{\Lambda _{11} \lr{\vec{u}}} \sum _{j = 1} ^k \Lambda_{jj} \lr{\vec{u}} \label{eq:flatten}
\end{equation} where $\vec{u}$ are the shaper's parameters. At this point we are not concerned with the absolute value of the gains, which can in principle be adjusted changing the power of the pump, so we divide all the gains by the largest one. For the second task, we run the optimization with the fitness function \begin{equation}
f_2 \lr{\vec{u}} = \frac{\Lambda _{11} \lr{\vec{u}}}{\Lambda _{22} \lr{\vec{u}}}. \label{eq:conc}
\end{equation} Note that since we are only interested in the gains, which are the singular values of the joint spectral distribution, and not the shape of the supermodes, we can use commonly available numerical routines for the SVD.\\
For the optimization to be meaningful some constraints have to be imposed. Indeed, if no constraint is imposed, the algorithm may converge to solutions which have a very small overlap with the Gaussian pulse that would be obtained without the shaper. This is a problem because, since the shaper is a passive optical component, it means that much of the power in the pulse is thrown away in the process and a very high power would be needed to realize such profiles. Optimization is however interesting because it makes clear that the "amount" of squeezing and its distribution among different modes are very different resources, as will be especially evident in the following sections about cluster states. More realistic profiles can be obtained with a modification to the fitness function which adds a weight hindering convergence towards profiles having a small overlap with the original Gaussian. To this end one can add a function of the power of the shaped pump, renormalized by the maximum of the shaper's transfer function to impose that the shaper is only attenuating~\footnote{Note that for numerical simulations we allow $\left|\mathcal{I}^{\lr{\vec{u}}} \lr{\omega} \right|>1$, hence the factor $1/m\lr{\vec{u}}^2$ }. The power of the shaped pump is given by \begin{equation} \label{eq:w}
w\lr{\vec{u}} = \frac{1}{m\lr{\vec{u}} ^2}\int \mathrm{d}\omega \left|\alpha^{\lr{\vec{u}}}\lr{ \omega } \right|^2
\end{equation} where \begin{equation}
m\lr{\vec{u}} = \max_\omega \left|\mathcal{I}^{\lr{\vec{u}}} \lr{\omega} \right|.
\end{equation}  The fitness functions $f_1$ and $f_2$ are then replaced by \begin{align}
\bar{f}_1 \lr{\vec{u}} &= \frac{1}{\Lambda _{11} \lr{\vec{u}}} \sum _{j = 1} ^k \Lambda_{jj} \lr{\vec{u}} + a \cdot x\lr{ w\lr{\vec{u}} } \label{eq:f1} \\
\bar{f}_2 \lr{\vec{u}} &=  \frac{\Lambda _{11} \lr{\vec{u}}}{\Lambda _{22} \lr{\vec{u}}} + b\cdot  y\lr{ w\lr{\vec{u}} }\label{eq:f2}
\end{align} with $a$ and $b$ positive real numbers. $x$ and $y$ may be arbitrary functions. A possible criterion to choose such a function may be that it should be negligible if the power is above some fraction of the original Gaussian and very rapidly becomes negative and large if the power is below this threshold. Solutions with a power lower than the threshold are then disfavoured but the weight does not influence the optimization as long as the power stays "acceptable". The magnitude of $a$ and $b$ can then be used for fine tuning. Fig.~\ref{fig:optSpectra} shows the results of two optimizations starting from a reference Gaussian spectrum with a shaper working in a window of about $9$~nm, corresponding to $\pm 3$ standard deviations (in amplitude) around the central frequency, and a $1.5$~mm BIBO crystal for down-conversion.\\
The supermodes resulting from the optimized pumps are shown in Fig.~\ref{subFig:supermodes}. At first sight both amplitude and phase seem very complicated, except for the first supermode arising from the optimization $\bar{f}_2$. This is a good sign, because the first supermode is the most interesting one in this case, being by far the most squeezed. As for the others, the apparent complexity may be explained and overcome by the quasi-degeneracy of the gains. See Appendix~\ref{app:supermodesComplexity} for details.\\
We stress that the optimization algorithm is stochastic and there is no guarantee that the optima are also global optima. Our aim here is to show that optimizing the shaper's cofiguration we could find pump profiles giving a significant improvement on the initial Gaussian. When $\bar{f}_1$ is optimized, the squeezing spectrum is made flatter, with $\Lambda_{jj} > 0.9 \Lambda_{11}$ for $j$ up to $\sim 80$, to be compared with $j \sim 30$ for a Gaussian pump. Optimizing $\bar{f}_2$ we found that a noticeable gap can be induced between the first and second gains, in this case $\Lambda_{11}/\Lambda_{22} \approx 1.43$, to be compared with $\Lambda_{11}/\Lambda_{22} \approx 1.00$ for a Gaussian pump. We stress that this quantity is essentially unrelated to the Schmidt number. Optimizing $f_2$ or $\bar{f}_2$ leads to a higer Schmidt number than that obtained with an unshaped pump. The reason is that many modes after the first one have almost the same squeezing. This kind of configuration could be interesting for example to work in a regime in which only the first supermode is pumped slightly above threshold. The classical noise of the pump laser would then be conveyed in the first supermode, improving the purity of the others, which would still be squeezed~\cite{aboveThr1, *aboveThr2}. The pump optimizing $ \bar{f}_1 $ carries about $30\%$ of the power of the unshaped Gaussian pump, while this figure is about $40\%$ for the pump optimizing $\bar{f}_2$, meaning they may realistically be implemented in the lab. The opportunity of introducing  the modifications in Eq.~(\ref{eq:f1}) and Eq.~(\ref{eq:f2}) is made more evident by comparison with the power of the pump profiles optimizing $f_1$ and $f_2$ (not shown here), which is of the order of $0.1\%$ of the unshaped Gaussian.\\
The procedure outlined in this section can be carried out for any function that can be written in terms of the shaper's parameters $\vec{u}$. For example maximizing the gain of the first supermode for a given maximum power, minimizing the spectral width of the first supermode, maximizing or minimizing the Schmidt number of the parametric gains as defined in~\cite{gatti, silber}, which gives a measure of the number of modes excited in the process. An example of interest for quantum information processing is treated in the following section.

\subsection{Cluster states on frexels}

We detail here how to optimize the profile of the pump to reduce the noise of the nullifiers of CV cluster states when the nodes of the graph are associated with a specific set of modes which have non-overlapping spectra.

\subsubsection{Definining the detection modes: frexels}

We turn our attention to a specific set of $m$ orthogonal modes which are slices of a Gaussian pulse. We refer to these as frexel modes (from "frequency elements") and denote their annihilation operators by $\left\{ \pi_j \right\}$. frexels can be seen as a specific realization of the modes $d$ in Eq.~(\ref{eq:dModes}). First, we choose a set of frequency bands of limits $  \lr{\Omega_1, \Omega_2}, ...,  \lr{\Omega_m, \Omega_{m+1} } $. The frexel modes are then defined by the spectral amplitudes \begin{equation}
 \left\{
\begin{array}{ll}
      \pi_j\lr{\omega} = \frac{e^{i\theta_j}}{\sqrt{\mathcal{N}_j}} \alpha^{\lr{\pi}}\lr{\omega} & \quad \Omega_j\leq \omega \leq \Omega_{j+1} \\
       \pi_j\lr{\omega} = 0 & \quad \mathrm{otherwise}
\end{array} 
\right.
\end{equation} where $\alpha^{\lr{\pi}}$ is a Gaussian pulse with a FWHM of $10$ nm centered around $2 \lambda_0 = 795$ nm, $\theta_j$ are arbitrary phases, which will turn out to be useful in the following, and~\footnote{If discrete frequency are considered the integrals are to be replaced by sums.} \begin{equation}
\mathcal{N}_j = \int \limits _{\Omega_j} ^{\Omega_{j+1}} \mathrm{d}\omega \left| \alpha^{\lr{\pi}} \lr{\omega} \right| ^2 .
\end{equation} An example with four frexels is depicted in Fig.~\ref{subFig:4frix}. The interest of these modes resides in the fact that, having non-overlapping spectra, they can be physically separated rather easily from one another using a prism or a grating~\footnote{A prism will not separarate $\pi_j$ from all the modes having the same support. Pixel modes as defined here make sense if one wishes to ultimately measure them through homodyne detection using a local oscillator shaped as $\alpha _{\mathrm{LO}}$. This is the simplest setting for CV information processing with multi-pixel homodyne detectors~\cite{ferrini2013compact}}. It is worth noting that, in principle, modes with an arbitrary spectral profile could be separated from a bunch of co-propagating modes~\cite{eckstein2011quantum, reddy2014efficient}, but this would involve nonlinear interactions which would make it unpractical to separate more than one mode from all the others. MBQC with frequency or spatial pixel modes was also introduced in~\cite{ferrini2013compact}. After being separated, frexels could be sent to different parties in a network or directly subject to independent homodyne measurements, for example. Indeed, the availability of multi-pixel homodyne detection schemes~\cite{armstrong2012programmable,armstrong2012programmable} is the main reason to introduce an overall Gaussian spectrum in the definition of frexel modes and an individual phase $\theta_j$ for each of them. The latter could be adjusted simply changing the phase of the local oscillator in each frequency band. This is an important degree of freedom to consider, as a phase shift of the local oscillator implies the measurement of a different quadrature, which is at the heart of CV-MBQC~\cite{gu2009universal}. Also, although a local phase-shift cannot change the amount of entanglement between frexels, it can change the kind of quantum correlations. As it will be clear from the next section, this is especially relevant for CV cluster states, which have a very specific type of correlations, resulting in the reduction of noise in the nullfier operators.

\subsubsection{Finding the optimal frexel permutation \label{sec:optPerm}}

\begin{figure}
    \subfloat[]{
        \centering
        \label{subFig:4frix}
        \includegraphics[width=0.24\textwidth]{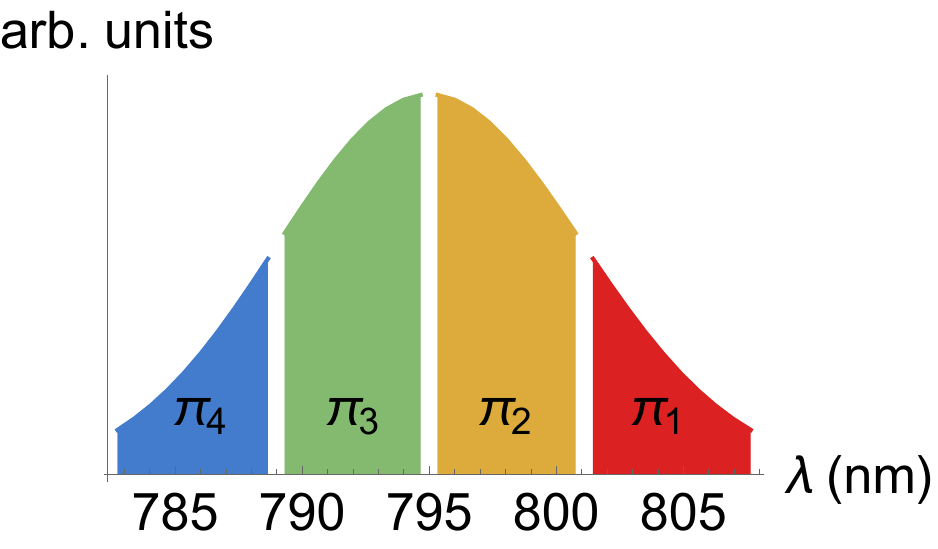}
    }
\subfloat[]{
        \centering
        \label{subFig:lin4modesClu}
        \includegraphics[width=0.24\textwidth]{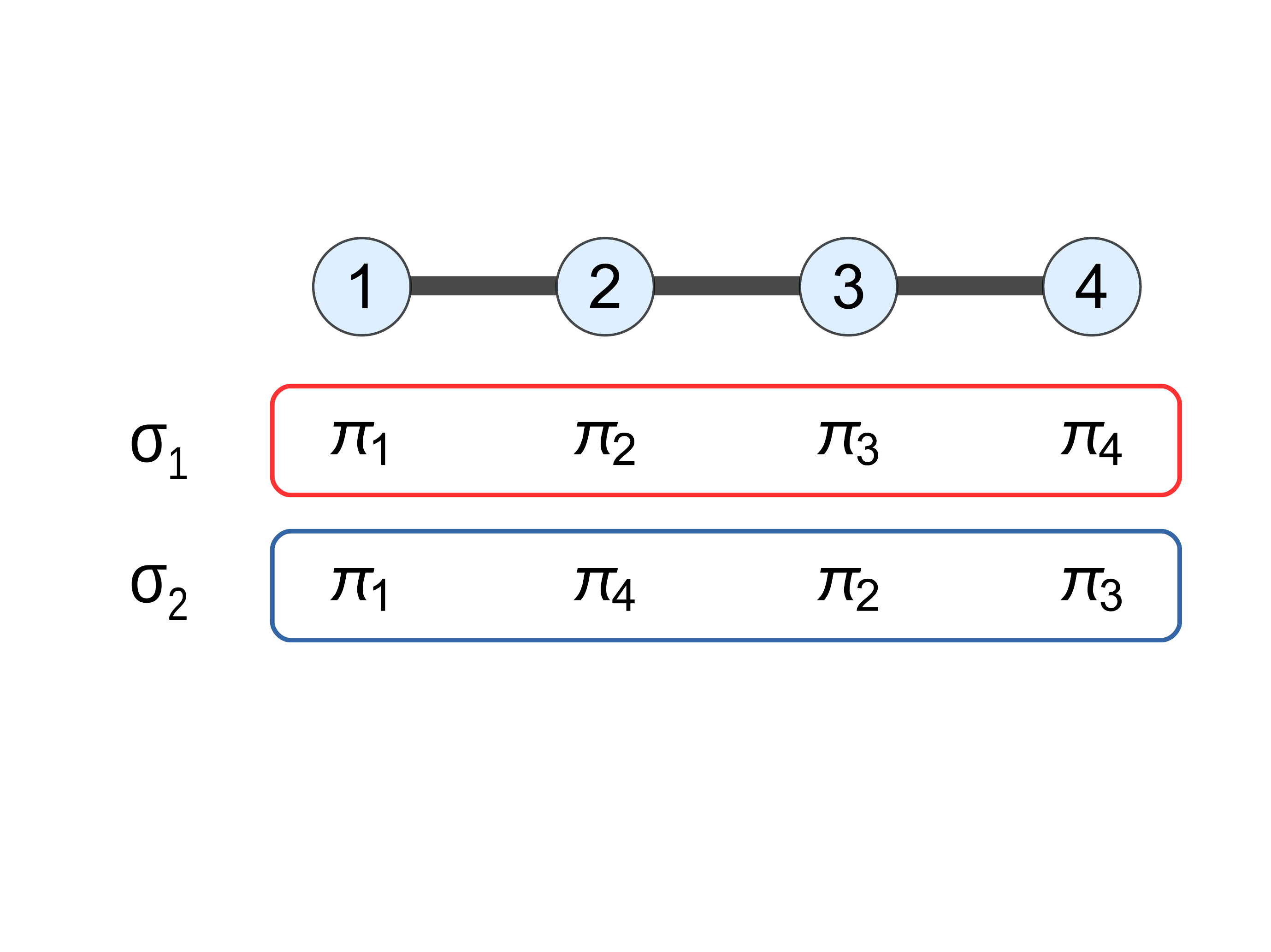}
    }  
    \caption{\label{fig:lin4modes}(Color online) (a) Spectral amplitude of four frexels within 3 standard deviations around the central frequency of the downconverted comb. The amplitudes are not normalized for clarity of representation. (b) A linear four-modes cluster state and two possible mappings of frexels onto its nodes. The second permutation $\sigma_2$ leads to smaller nullifiers' noise for an appropriate choice of the global phase of each pixel (not shown in the drawing). } 
\end{figure}

We consider, as an example, four frexels, and associate $\pi_j$ to the $j$th node of the 4-mode linear cluster state depicted in Fig.~\ref{subFig:lin4modesClu}, corresponding to the graph with adjacency matrix \begin{equation}
G_{\mathrm{lin}} = \lr{ \ba{cccc} 0 & 1 & 0 & 0 \\ 1 & 0 & 1 & 0 \\ 0 & 1 & 0 & 1 \\ 0 & 0 & 1 & 0\ea}.
\end{equation} This cluster state is universal for single-mode Gaussian MBQC~\cite{lubo}. We can compute the variance of nullifiers using the procedure explained in Sec.~\ref{sec:nulls} for a general set of modes. The choice of the local phases $\theta_j$ defines which quadratures correspond to amplitude $\vec{q}_\pi$ and which to phase $\vec{p}_\pi$. Since we assumed to be free to choose an independent phase reference for each pixel, we can use the $\theta_j$ giving the lowest fluctuations for the nullifiers on average. For the numerical calculation, we assume that the unshaped pump is a Gaussian of amplitude $\alpha ^{\lr{g}}$  (see Eq.~(\ref{eq:refGauss})) and that down conversion happens in a $0.5$ mm BIBO crystal. We fix the pump power so that the squeezing in the leading supermode is $7$ dB. We take $2\pi c / \Omega_1 \simeq 808$~nm and $2\pi c / \Omega_5 \simeq 782$~nm. The average of nullifiers' variances is found to be $\Delta^2 _\mathrm{avg} \delta \simeq 0.49$ (vacuum is normalized to $0.5$), which amounts to a noise reduction of about $-0.08$ dB. The same calculation may be carried out for any permutation $\sigma$ of frexels, namely assigning $\pi_{\sigma\lr{j}}$ to node $j$ on the graph. It turns out that some permutations allow to sensibly reduce the average noise of nullifiers. For example $\Delta^2 _\mathrm{avg} \delta \simeq 0.29$ for the permutation $\sigma_2 = \lr{\pi_1,\pi_4,\pi_2,\pi_3}$, corresponding to about $-2.35$ dB. This may look surprising at first, since a simple relabeling of the modes cannot change the amount of entanglement. The point is that nullifiers' noise reduction is not just a signature of entanglement, but rather of very specific correlations among the nodes of the corresponding graph, and these may very well vary from one permutation to the other. In our example, the linear graph has a link between nodes $1$ and $2$, corresponding to frexels $\pi_1$ and $\pi_2$ if the trivial permutation is considered and to frexels $\pi_1$ and $\pi_4$ if one instead considers the $\sigma_2$. Being symmetric with respect to the central frequency, we expect frexels $\pi_1$ and $\pi_4$ to be more entangled after the downconversion than frexels $\pi_1$ and $\pi_2$ whose spectra are on the same side of the central frequency of the downconverted field. We then expect a better noise reduction in the corresponding nullifier. The permutation $\sigma_2$ is actually the optimal for the conditions considered here.

\subsubsection{Optimal pump profiles}

\begin{figure}

\subfloat[]{
        \centering
        \label{subFig:lin4modesNulls}
        
        \includegraphics[width=0.37\textwidth]{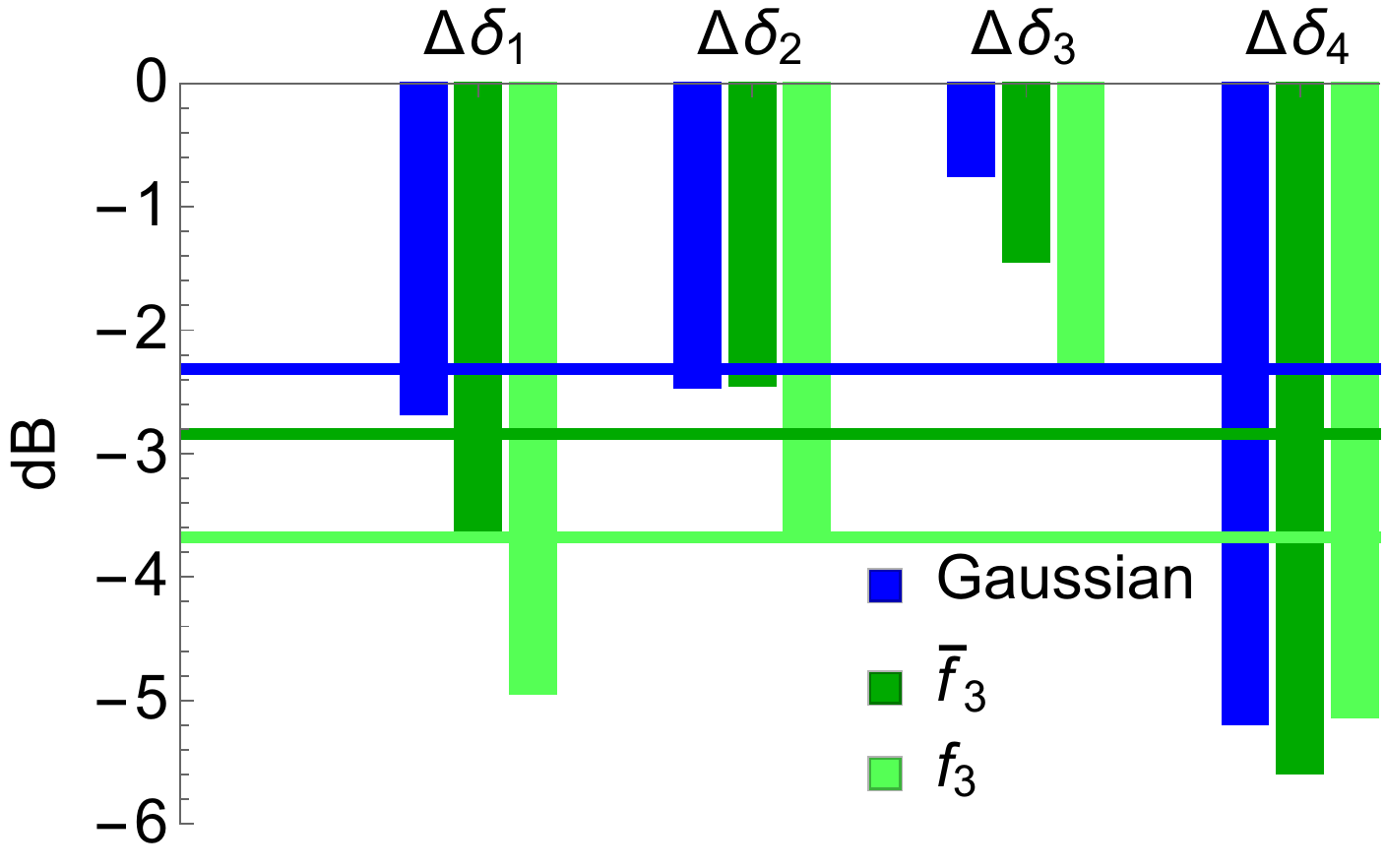}
    }  \\
\subfloat[]{
        \centering
        \label{subFig:lin4modesPumpNoW}
        
        \includegraphics[width=0.45\textwidth]{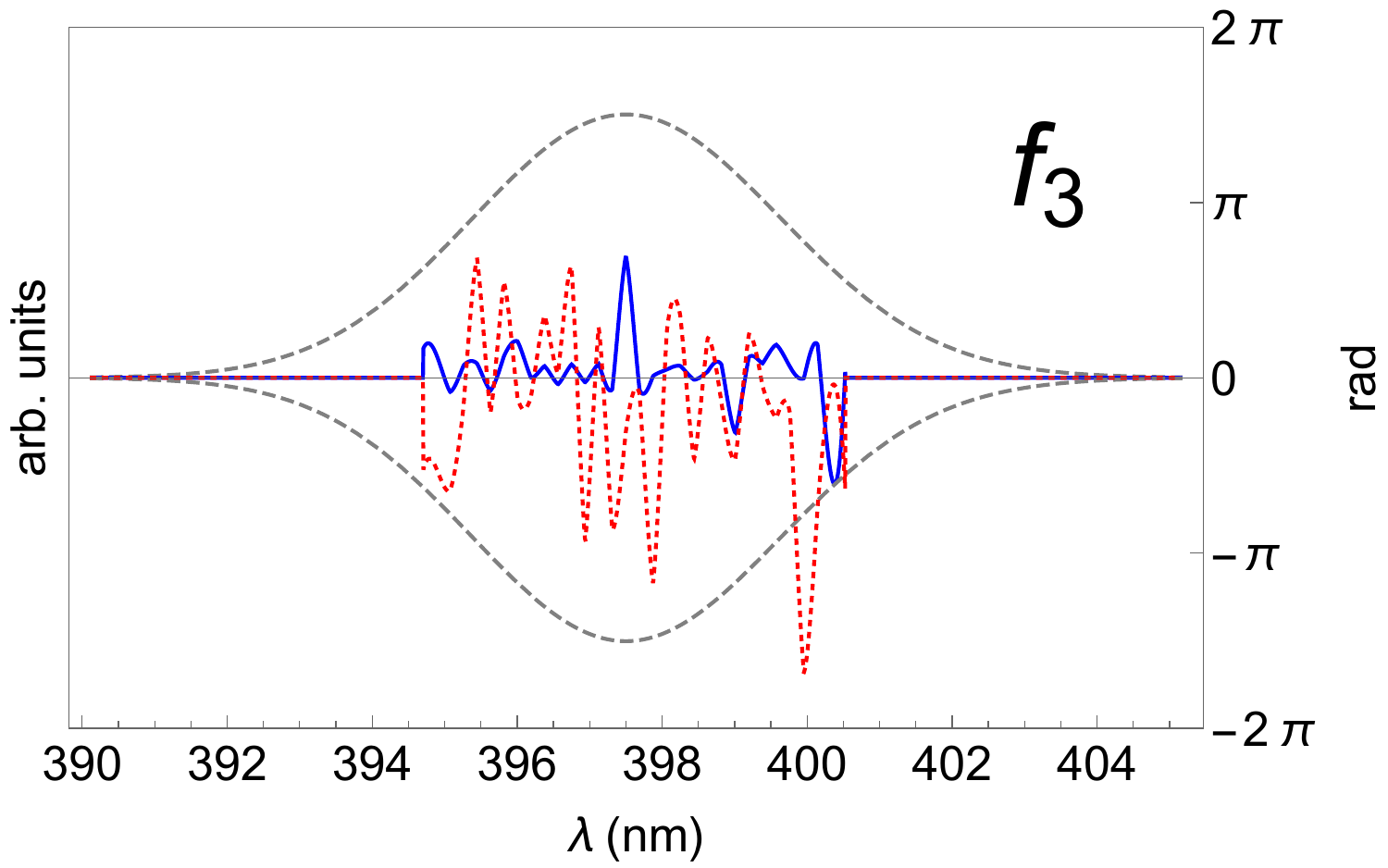}
    }   \\ 
\subfloat[]{
        \centering
        \label{subFig:lin4modesPumpPW}
        
        \includegraphics[width=0.45\textwidth]{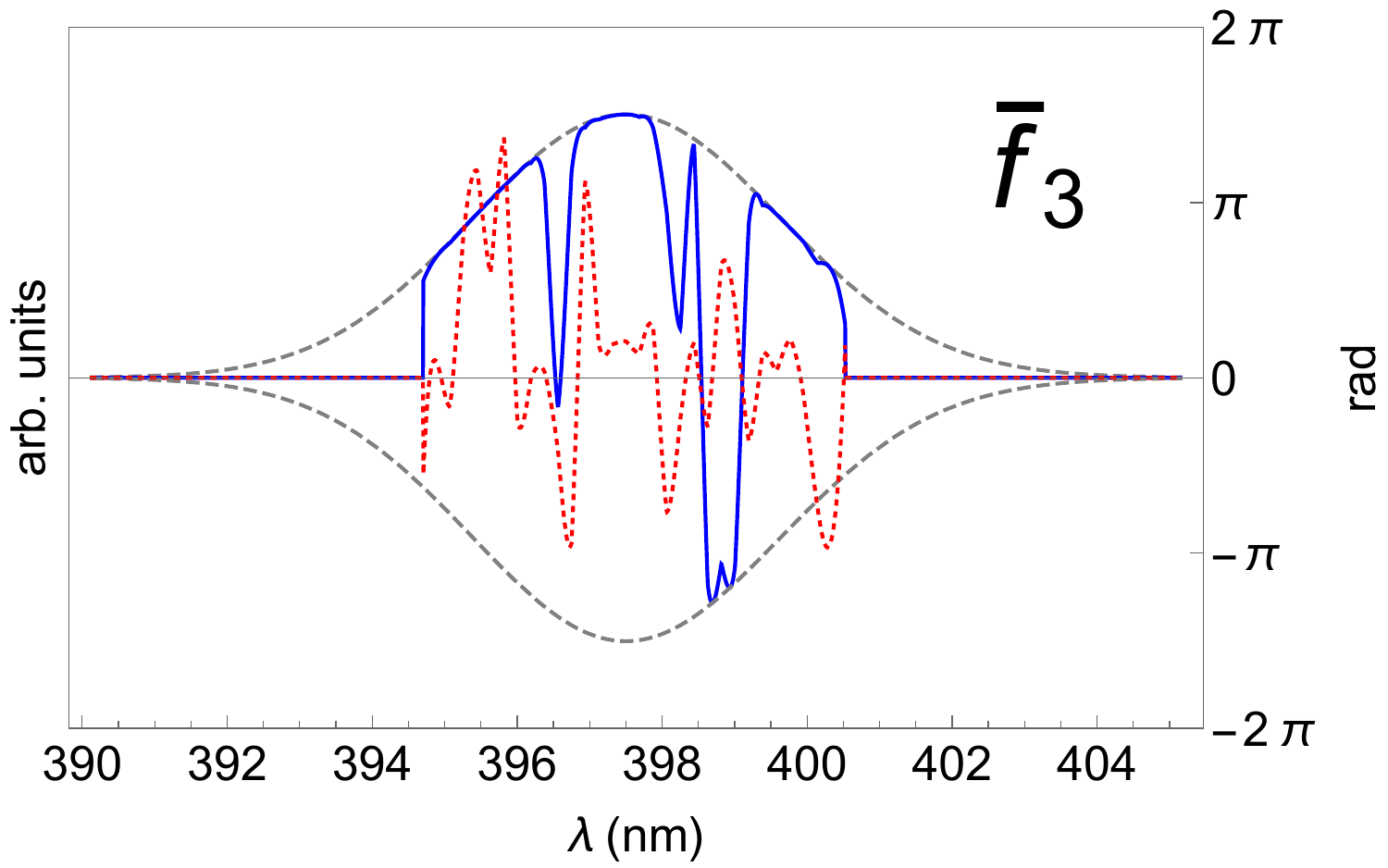}
    }         
    \caption{\label{fig:lin4CluOpt}(Color online) Results of the optimization of the pump shape to reduce the average noise of the nullifiers of a four-modes linear cluster.  (a) shows the nullifiers' noise reduction in dB for a Gaussian pump and for the optimal profiles found optimizing $f_3$ (Eq.~(\ref{eq:f3})) and $\bar{f}_3$ (Eq.~(\ref{eq:f3bar})) with $c = 1.35 $. The squeezing of the leading supermode was fixed to $7$ dB. The bar on the left of each triplet (dark blue) corresponds to the Gaussian case, the central bar (dark green) to $\bar{f}_3$, the bar on the right (light green) to $f_3$. The horizontal lines show the average squeezing in each case (The Gaussian case, $f_3$ and $\bar{f}_3$ correspond to the top, middle and bottom lines, respectively).  The pump profiles optimizing $f_3$ and $\bar{f}_3$ are shown in (b) and (c) respectively. The scale on the left refers to  amplitude, while that for the phase is on the right.}
\end{figure}

Starting from the best permutation in the previous section, we used numerical optimization to find the pump profiles minimizing the function~\footnote{$\Gamma_{\bar{\delta}\bar{\delta}}$ is a covariance matrix, so it is positive-semidefinite by construction. As a consequence $\mathrm{Tr}\left[\Gamma_{\bar{\delta} \bar{\delta}} \right] \to 0$ is equivalent to $\Gamma_{\bar{\delta} \bar{\delta}} \to 0$} \begin{equation}\label{eq:f3}
f_3\lr{\vec{u}}  = \mathrm{Tr}\left[\Gamma_{\bar{\delta} \bar{\delta}} \lr{\vec{u}} \right]
\end{equation} with $\Gamma_{\bar{\delta} \bar{\delta}}$ defined as in Eq.~(\ref{eq:nullsCov}) for the four-modes linear cluster. For the optimization, we start from a reference Gaussian pump and assume the shaper is acting on a spectral window of $ \pm 2 $ standard deviations around the central frequency, corresponding to approximately $ 95\% $ of the pump power. First, we fixed the squeezing of the leading supermode to $7$~dB, which is consistent with the highest values measured in experiments with frequency combs~\cite{renne}. Analogously to the case of the squeezing spectrum, the algorithm converges to pump profiles which have a small overlap with the original pulse, so we also ran the optimization for the modified function \begin{equation}\label{eq:f3bar}
\bar{f}_3\lr{\vec{u}}  = \mathrm{Tr}\left[\Gamma_{\bar{\delta} \bar{\delta}} \lr{\vec{u}} \right] - h \cdot w\lr{\vec{u}}
\end{equation} where $h$ is a positive real number and $w$ is defined as in Eq.~(\ref{eq:w}). The results are shown in Fig.~\ref{fig:lin4CluOpt}. Optimization of $f_3$ leads to a larger improvement of the nullifiers squeezing on average, but as shown in Fig.~\ref{subFig:lin4modesPumpNoW} the corresponding pump profile has a small overlap with the original Gaussian. As a consequence, the shaped pulse only contains $\sim 2\%$ of the power of the unshaped pulse. Optimization of $\bar{f}_3$ leads to a profile (Fig.~\ref{subFig:lin4modesPumpPW}) that still allows to reduce the average nullifiers' noise of about $0.5$ dB with respect to the Gaussian profile while containing $\sim 80\%$ of the Gaussian pulse's power. This could lead to a measurable improvement in realistic experimental conditions. The compromise between power in the shaped pump and noise reduction can be tuned changing the parameter $h$ in Eq.~(\ref{eq:f3bar}) in order to adapt to specific experimental constraints. If more power is available, for example, the optimization could be performed for smaller values of $h$.

\subsubsection{Relation between highest squeezing and nullifiers' noise}

\begin{figure}
\vspace{0.3cm}
    \subfloat[]{        
        \centering
        \label{subfig:sqzDepNullsAv7dB}
        \includegraphics[width=0.4\textwidth]{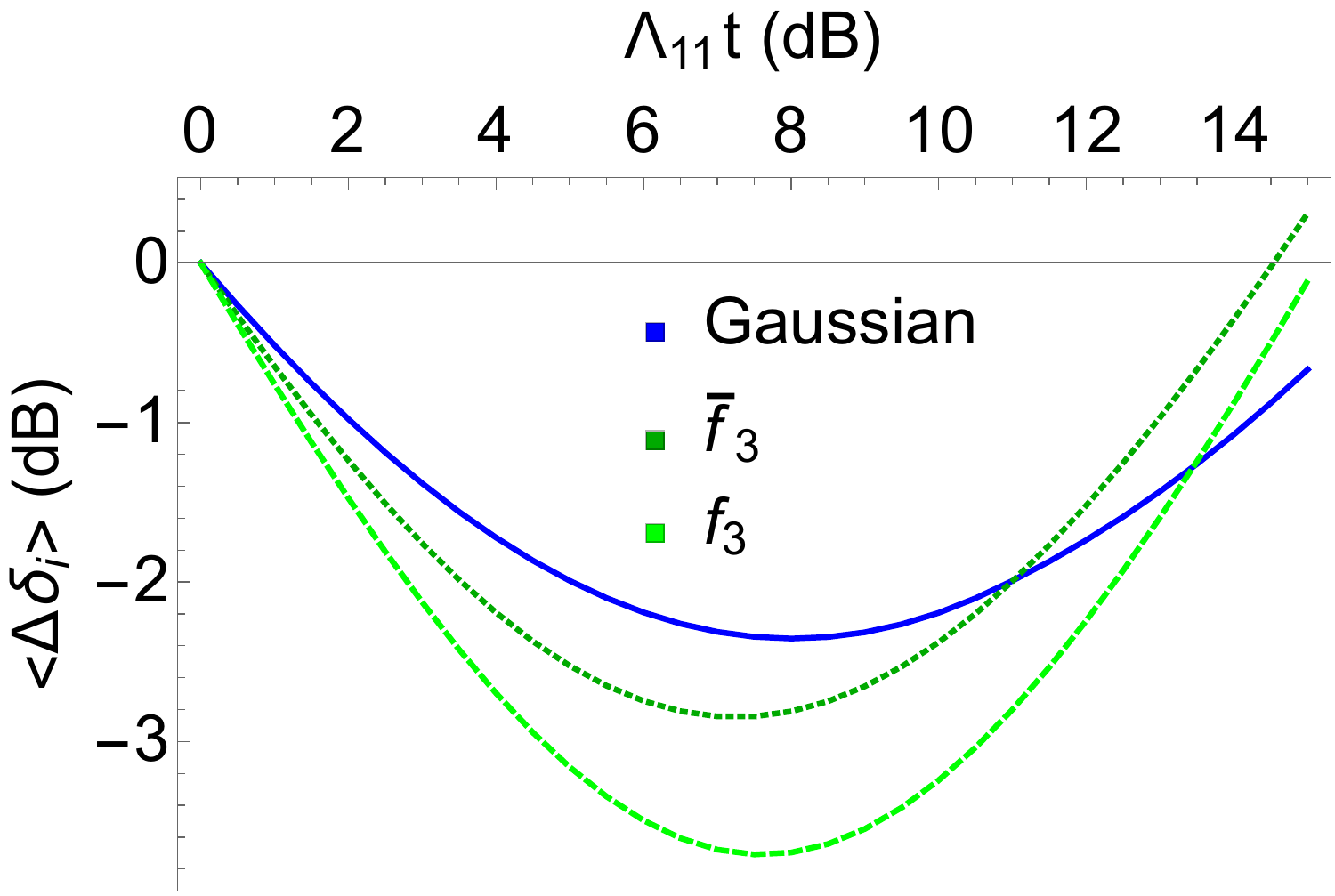}
    }\\
    \subfloat[]{ 
        \centering
        \label{subfig:sqzDepNullsAv20dB}
        \includegraphics[width=0.4\textwidth]{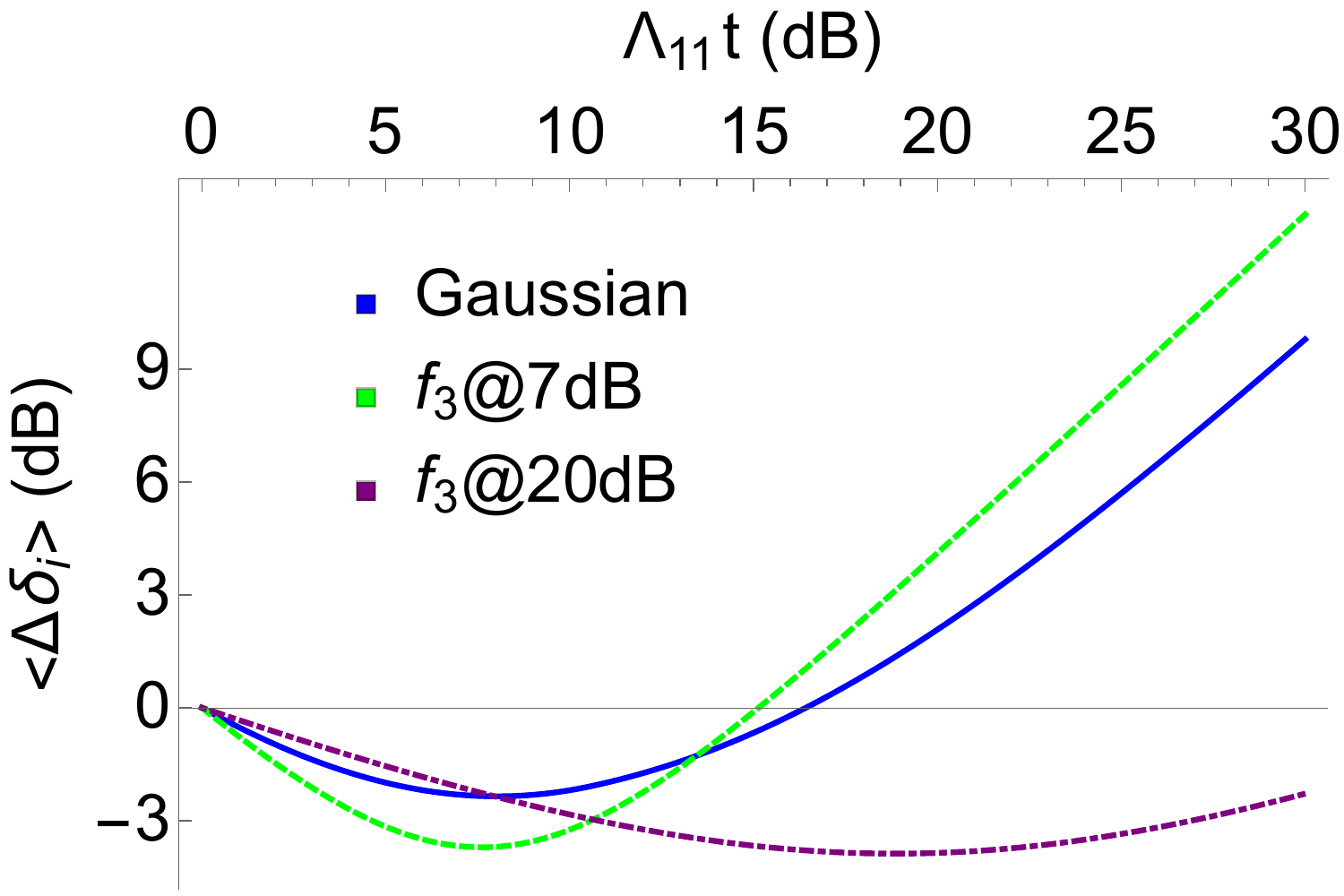}
    }
    \caption{\label{fig:sqzDepNullsAv}(Color online) Average nullifier's noise for a linear cluster on four frexel modes as a function of the squeezing of the leading supermode. The curves in~(\ref{subfig:sqzDepNullsAv7dB}) are obtained for a Gaussian pump (solid blue line) and the pump profiles obtained optimizing $f_3$ (dashed light green line) and $\bar{f}_3$ (dotted dark green line) fixing the leading supermode's squeezing to $7$ dB, while in~(\ref{subfig:sqzDepNullsAv20dB}) the curves for a Gaussian pump (solid blue line) and the configurations optimizing $f_3$ for a squeezing of $7$~dB (dashed light green line) and $20$~dB (dot-dashed purple line) of the leading supermode are shown.} 
\end{figure}

It is interesting to compute what happens when one changes the pump power keeping the shaper's configuration fixed. As long as the low-gain or below-threshold conditions are satisfied, this should just multiply the gains by a common factor. One could try and guess that more power, meaning a higher squeezing in all supermodes, would imply better noise reduction for the nullifiers. This is not actually the case, as can be seen from Fig.~\ref{fig:sqzDepNullsAv}. In fact, the average nullifiers' noise is reduced from the shot noise  until a certain value of the squeezing of the first supermode. If the power of the pump is further increased, the average nullifiers' noise starts increasing as well. One explanation could be that the number of squeezed modes in the system largely exceeds the number of frexels, so the contribution of all anti-squeezed quadratures to the nullifiers cannot be made arbitrarily small. The optimal configuration is found minimizing the contribution of the leading anti-squeezed quadratures. But even if the remaining anti-squeezed quadratures appear in the nullifiers with very small coefficients, at some point the corresponding noise will dominate, since it grows indefinitely with the gain.

Running the optimization with the squeezing of the leading supermode set to a different value results in a different optimal pump profile. With this different profile, the average nullifiers' noise will attain a minimum when the squeezing of the first supermode is close to the one chosen for the optimization. An example is shown in Fig.~\ref{subfig:sqzDepNullsAv20dB}, where the average nullifiers' noise as a function of the leading squeezing for a Gaussian pump and two profiles optimized at different leading squeezing are compared.

\section{Conclusions \label{sec:conclusions}}

In summary, we showed that pump shaping can be used effectively to engineer the quantum state produced by the spontaneous parametric down-conversion of a frequency comb. To this end, we introduced a method, based on either Autonne-Takagi or, equivalently, Bloch-Messiah decomposition, that can be applied to the numerical study of any spectral profile, including pulses with a general frequency-dependent phase. As a first example, we used this method to study the effect of spectral chirping, which is commonly met in experiments with frequency combs, on the down conversion process. We found that the quadratic phase has a non trivial effect on the parametric gains and on the spectral profile of the supermodes. Furthermore, using an optimization algorithm we found optimal profiles for flattening the values of the parametric gains or creating a gap between the gain of the first and second supermodes. In both cases we showed that the shape of the pump has a macroscopic effect on the output state which can lead to measurable improvements in realistic experimental conditions. We then applied the same technique to find the pump profiles which are optimal to produce CV cluster states when the nodes of the cluster correspond to spectral slices of a Gaussian pulse. We focused on a four-mode linear cluster state. This is universal for single-mode Gaussian CV-MBQC, so our results are directly applicable to CV-MBQC with frexel modes. Similar results can be obtained for different graphs, such as the six-mode centered pentagon used for CV secret sharing protocols in~\cite{van2011implementing} and~\cite{cai2017multimode}.

We stress that our approach is very general and, besides the examples cited here, it can be applied with small modifications to optimize any property of the output state after the down-conversion, such as the squeezing of the leading supermode or the Schmidt number. The same approach was used, for example, in a recent work proposing the simulation of quantum complex networks with an all-optical setup~\cite{networks}.

Finally, we note that our results rely on the use of a non-deterministic optimization routine. Our goal was to show the effectiveness of the overall approach but we did not compare the performances of this specific algorithm with others. On the other hand, the general procedure is the same if a different routine is used. The results may then potentially be improved using a different optimization algorithm. Also, conceptually the same approach can be used in closed-loop experiments in which the fitness function is replaced by a measured quantity.

\section*{Acknowledgments}

We thank Valentina Parigi, Adrien Dufour, Giuseppe Patera, Valerian Thiel, Dmitri Horoshko and Mikhail Kolobov for useful discussions. We thank Tobias Lipfert for interesting discussions and for pointing out ref.~\cite{takagiSVD} to us. We are grateful to the anonymous referee for the careful review and for providing constructive inputs on earlier versions of the manuscript.
This work is supported by the European Union Grant QCUMbER (No. 665148) and the French National Research Agency projects COMB and SPOCQ. N. T. is a member of the Institut Universitaire de France.

\appendix

\section{Phase-matching SPDC in BiBO}
\label{app:phiMBiBO}

This appendix briefly reviews how phase matching is achieved for SPDC in  $\mathrm{BiB}_3\mathrm{O}_6$ crystals, commonly known as BiBO \cite{Ghotbi}. In practice birefringence is exploited to match the propagation velocity of pump and signal/idler fields.

BiBO is a biaxial crystal. The dispersion relations for polarized light propagating along one of the axis $x$, $y$ or $z$ can be computed using Sellmeier's equations \begin{equation}
n_i \lr{ \lambda } = \sqrt{ A_i + \frac{B_i}{\lambda^2 - C_i} - D_i \lambda ^2 }
\end{equation} where $i = x,\ y,\ z$ and $\lambda$ is the wavelength. The Sellmeier's coefficients are \\

\begin{center}
\begin{tabular}{ccccc}
Index & $A_i$ & $B_i$ & $C_i$ & $D_i$ \\
\hline \\
$n_x$ & 3.07403 & 0.03231 & 0.03163 & 0.013376 \\
$n_y$ & 3.16940 & 0.03717 & 0.03483 & 0.01827 \\
$n_z$ & 3.6545 & 0.05112 & 0.03713 & 0.02261 \\

\end{tabular}
\end{center}

Consider a plane wave of wave vector $\boldsymbol{k}$ propagating in the medium. We denote by $\Pi$ the plane perpendicular to $\boldsymbol{k}$ and containing the origin of the ellipsoid $\mathcal{E}$ of indices. For historical reasons, the phase mathing angles $\theta$ and $\phi$ describing the rotation of $\mathcal{E}$ with respect to its axes is described with geographical coordinates, so the triad of axis is left-handed. $\phi$ is the angle from the $xz$ plane to the $yz$ plane and $\theta$ is the angle from $y$ to $z$. The refractive index for given wavelength and propagation direction is determined through \begin{equation} \label{eq:refIndGenSell}
\frac{1}{n\lr{\lambda, \theta, \phi} } = \sqrt{\frac{\cos^2\theta\cos^2\phi}{n^2 _x \lr{\lambda}} + \frac{\cos^2\theta\sin^2\phi}{n^2 _y \lr{\lambda}} + \frac{\sin^2\theta}{n^2 _z \lr{\lambda}}} .
\end{equation}
According to \cite{Ghotbi}, BiBO can phase-match Type I ($e +e \to o$) processes with $\phi = \pi /2$ for signal and idler and $\theta$ varying depending on the fundamental wavelength. For SPDC, this means that we can take the pump field polarized along $x$ ($\theta = 0$) and the polarization of signal and idler in the $yz$ plane. Eq.~(\ref{eq:refIndGenSell}) gives for the the refraction index of signal and idler \begin{equation}
n_e \lr{\lambda, \theta} = \lr{ \frac{\cos^2\theta}{n^2 _y \lr{\lambda}} + \frac{\sin^2\theta}{n^2 _z \lr{\lambda}}  }^{-\frac{1}{2}}.
\end{equation} We consider a collinear configuration and denote by $2\omega_0$ the central frequency of the pump. The down-converted field will then be centered around $\omega_0$. The phase matching condition requires that the phase mistmatch (Eq.~(\ref{eq:phaseMismatch})) is zero for the central frequencies \begin{equation} \label{eq:phiMCond}
k_p \lr{2\omega_0 } - 2 k_e \lr{  \omega _0  ,\theta} = 0
\end{equation} with \begin{align}
k_p \lr{\omega} = \frac{\omega n_x \lr{\frac{2\pi c}{\omega}}}{c}\\
k_e \lr{\omega,\theta} = \frac{\omega n_e \lr{\frac{2\pi c}{\omega}, \theta}}{c}.
\end{align} Eq.~(\ref{eq:phiMCond}) is then satisfied if $n_e \lr{2\pi c /\omega_0, \theta} = n_x \lr{\pi c /\omega_0}$. Assuming that the central wavelength of the pump is $2 \pi c /2\omega_0 = 397.5$ nm, this is achieved for $\theta = 2.63214$ ($\theta = 150.811^\circ$).

\section{Details of the numerical simulation \label{app:numDet}}

We are mainly concerned with optical frequency combs, in which case the number of frequency modes involved is of the order of $~10^5$. Using the full comb to describe the system would make the problem numerically intractable. We adopt then a coarse-grained description of the system, treating first the comb as a continuum and then discretizing the problem. This may also be motivated by the fact that the free spectral range is too small for the single teeth of the comb to be resolved in the experiments. Moreover, the average photon number for the single frequency is too low to display quantum features, each frequency mode being essentially in the vacuum state. We took about $~500$ points for the discretization, which is close to the number of the physical pixels of commercially available pulse shapers. Obviously, ideally the state is mixed in this coarse grained desription, but our approximation turns out to be very good as long as the number of frequencies we take into account is large enough to represent all the supermodes which are significantly squeezed. Throughout this work, frequency modes will be identified with the coarse grained frequency pixels, although analytical calculations rigorously hold only for the teeth of the comb.

\section{Model of the pulse shaper \label{sec:shaper}}

The shaper is modeled as a function interpolating the values $\vec{u}$ of amplitude and phase at 32 frequencies within a spectral window centered at the central frequency of the Gaussian pump comb $\omega _0$. For our calculations we chose the half width of the window to be two or three times the standard deviation of the Gaussian. The Gaussian comb is then multiplied by the function $\mathcal{I} ^{\lr{\vec{u}}}  \lr{\omega}$ to obtain the shaped pump. The number of frequencies independently controlled by the shaper is taken to be between 20 and 40, which is compatible with the spectral resolution of the shaper in a 4-f configuration~\cite{pulseShaper}. As a consequence, the complex amplitude of the pump pulse after shaping, as well as the spectrum of the supermodes and the respective gains, will depend on the vector of real parameters $\vec{u}$. Interpolation is needed to smooth the solution for the output comb and obtain pulse shapes which could be practically realized in experiments. In fact, despite the large number of pixels in the shaper, the configuration of neighbouring pixels is correlated due to electromagnetic interactions, which makes, for example, a $\pi$ phase between neighbouring pixels practically impossible to realize. We chose to interpolate the function with cubic polynomials. 

\section{Optimization algorithm \label{app:opt}}

The algorithm mimicks Darwinian evolution to stochastically explore the parameter space and uses statistical analysis to find the direction of fastest ascent of a fitness function. It goes as follows: first, a point in the parameter space is chosen at random. A new generation, that is a number of mutations (which grows logarithmically with the dimension of the parameter space) is generated around the first point. At the first iteration, the mutations are generated according to an isotropic Gaussian distribution. The fitness function is evaluated for each mutant. The best half of the mutants are linearly combined to generate a new starting point for the algorithm. Since the algorithm was initially developed for applications in experiments, to mitigate the effect of experimental noise the new point is actually a combination of the mutants and the starting points of previous generations. Statistical analysis is then performed on the current generation to find the axes corresponding to greater improvement of the fitness function. The covariance matrix for the next generation is modified accordingly, stretching the corresponding axes. A general step size parameter is also adjusted as follows: if the direction of fastest ascent was roughly the same in the last generations, then the algorithm is travelling in the good direction and the step-size is increased. If the direction changed many times over the last generations, the algorithm is probably close to an optimum and the step-size is decreased to accelerate convergence.

In our case the parameters of the optimization will be the vector $\vec{u}$ of amplitude and phase parameters of the shaper introduced in Sec.~\ref{sec:numDet} and the fitness functions will be derived from properties of the output state obtained pumping the down-conversion process with the corresponding shaped pulse.

\section{Nullifiers's fluctuations through homodyne detection \label{app:nullsMeas}}

In fact, even though each $\delta_j$ in Eq.~(\ref{eq:nulls}) is not the quadrature of a mode, its normalized version is. Let us define $\bar{\delta}_j \equiv r_j \delta_j $ where $r_j$ is a real number such that \begin{equation}
\Delta ^2 _0 \bar{\delta}_j = \frac{1}{2}
\end{equation} if the field is in the vacuum state. Then it is possible to find a mode whose amplitude quadrature is precisely $\bar{\delta}_j$.  The normalization $r_j$ is readily computed as $r_j = 1 / \sqrt{1 + N\lr{j}}$, with $N\lr{j}$ the number of nearest neighbours of node $j$.

Using the definition of quadratures for the $\vec{d}$ modes $d_j = \lr{ q^{\lr{d}}_{j} + i p^{\lr{d}}_{j} } /\sqrt{2}$ and Eq.~(\ref{eq:dModes}), $\bar{\delta}_j$ may be rewritten as \begin{equation}
\bar{\delta}_j = \frac{1}{\sqrt{2} } \lr{ \sum _l W_{jl} a_l + \sum _l W_{jl} ^* a_l ^\dagger } \equiv \frac{1}{\sqrt{2} } \lr{A_j + A_j ^\dagger }
\end{equation} where $A_j$ is the annihilation operator associated with the mode defined by the spectral amplitudes \begin{equation} \label{eq:Wampl}
W_{jl} =  - r_j\lr{ iD_{jl} + \sum _k G_{jk}D_{kl} }.
\end{equation} These are the amplitudes of the electic field to print on the local oscillator in order to measure $\bar{\delta}_j$. They may as well be used to define a transformation $R_W$ analogous to $R_D$ in Eq.~(\ref{eq:RV}). Accordingly, one finds the covariance matrix associated with the nullifiers, which contains their squeezing as well as correlations between them and the conjugated operators $\bar{\zeta}_j$ \begin{equation} \label{eq:nullsCov}
\Gamma_{\bar{\delta}} = R_W \Gamma_\omega R_W ^T =  \lr{ \ba{cc} \Gamma_{\bar{\delta} \bar{\delta}} & \Gamma_{\bar{\delta} \bar{\zeta} } \\ \Gamma_{\bar{\delta} \bar{\zeta}  } ^T &   \Gamma_{\bar{\zeta} \bar{\zeta}} \ea }.
\end{equation} For an ideal cluster state $\Gamma_{\bar{\delta} \bar{\delta}} \to 0$~\cite{graphCalc}. Note that $\Gamma_{\bar{\delta}}$ is a legitimate covariance matrix, in the sense that it contains variances and covariances of the normalized nullifier operators, even if the corresponding modes, defined by the rows of $W$ in Eq.~(\ref{eq:Wampl}), are not always orthogonal.

\section{Linear combinations of quasi-degenerate supermodes \label{app:supermodesComplexity}}

Consider the first supermodes resulting from the optimization of $\bar{f}_1$ and the associated gains $ \lambda _j \equiv \Lambda_{jj}$. Since $\lambda_{20} > 0.99 \lambda_{1}$ and $\lambda_{30} > 0.97 \lambda_{1}$, when the first supermode has $5$ dB of squeezing, the difference of squeezing with the thirtieth supermode is about $\approx 0.13$ dB, while the squeezing of the twentieth supermode differs by less than $0.05$ dB from that of the first. This difference would hardly be detectable in experiments. As a consequence, it is reasonable to ask whether a linear combination of the first supermodes could by approximated by a simpler shape. Note that the coefficients in these linear combinations need to be real if one wants the resulting mode to be squeezed. One finds that a real Gaussian mode of about 37 nm FWHM has more than $92  \%$ overlap with a real combination of the $30$ first supermodes. If one allows for a linear phase, which only amounts to a delay, as we noted earlier, the overlap is about $98$\% for a Gaussian amplitude of about 24 nm FWHM considering a real combination of $22$ supermodes. The state of the modes resulting from these superpositions of supermodes would be slighly inpure but would nonetheless display squeezing in one quadrature which would exceed that of the least squeezed supermode in the linear combination.


\bibliography{./pumpShaping4}{}
\addcontentsline{toc}{chapter}{\bibname}

\end{document}